\documentclass[lettersize,journal]{IEEEtran}

\usepackage{amsmath,amssymb,amsfonts, bm}
\usepackage{algorithmic}
\usepackage{algorithm}
\usepackage{array}
\usepackage{kotex}
\usepackage{soul}

\usepackage[caption=false,font=normalsize,labelfont=sf,textfont=sf]{subfig}
\usepackage{textcomp}
\usepackage{stfloats}
\usepackage{url}
\usepackage{verbatim}
\usepackage{graphicx}
\usepackage{cite}
\usepackage{multirow}
\usepackage{enumerate}
\usepackage{xcolor}
\def\BibTeX{{\rm B\kern-.05em{\sc i\kern-.025em b}\kern-.08em
    T\kern-.1667em\lower.7ex\hbox{E}\kern-.125emX}}
\usepackage{balance}
\usepackage[flushleft]{threeparttable}
\usepackage{tabularx}
\newcommand{\meq}[2]{\begin{equation}\begin{alignedat}{3}#1\end{alignedat}\label{#2}\end{equation}}

\def\where{\text{ where }}
     
 \def\bf{{\mathbf{f}}}  
     
     \def\bp{{\mathbf{p}}}
    \def\bs{{\mathbf{s}}} 
  \def\bv{{\mathbf{v}}}   
\def\by{{\mathbf{y}}}  
   
 \def\bF{{\mathbf{F}}}

 \def\bV{{\mathbf{V}}}  \def\bX{{\mathbf{X}}}
\def\bY{{\mathbf{Y}}}

\def\T{^\mathsf{T}}

\newcommand{\rev}[1]{{\textcolor{black}{#1}}}

\newcommand*\xbar[1]{%
  \hbox{%
    \kern 0.1em
    \vbox{%
      \hrule height 0.5pt 
      \kern0.3ex
      \hbox{%
        \kern-0.0em
        \ensuremath{#1}%
        \kern-0.0em
      }%
    }%
    \kern 0.0em
  }
} 

\begin{document}
\bstctlcite{IEEEexample:BSTcontrol}

\title{EchoScan: Scanning Complex Room Geometries via Acoustic Echoes}

\author{Inmo Yeon, \IEEEmembership{Student Member, IEEE}, Iljoo Jeong, Seungchul Lee, and Jung-Woo Choi, \IEEEmembership{Member, IEEE}
\thanks{Manuscript received February 23, 2024}

\thanks{This work was supported by the National Research Foundation of Korea (NRF) grant funded by the Ministry of Science and ICT of Korea government (MSIT) (No. RS-2024-00337945), the BK21 FOUR program through the NRF grant funded by the Ministry of Education of Korea government (MOE), the Institute of Civil-Military Technology Cooperation funded by the Defense Acquisition Program Administration and Ministry of Trade, Industry and Energy (MOTIE) of the Korean government (No. 19-CM-GU-01), and the Korea Institute of Energy Technology Evaluation and Planning (KETEP) grant funded by the MOTIE of the Korean government (No. 20206610100290).

\textit{Inmo Yeon and Iljoo Jeong equally contributed to this work. (Corresponding authors: Seungchul Lee and Jung-Woo Choi.)
}}
\thanks{Inmo Yeon and Jung-Woo Choi are with the School of Electrical Engineering, Korea Advanced Institute of Science and Technology (KAIST), Daejeon, 34141, South Korea. (e-mail: iyeon@kaist.ac.kr; jwoo@kaist.ac.kr)}
\thanks{Seungchul Lee is with the Department of Mechanical Engineering, Korea Advanced Institute of Science and Technology (KAIST), Daejeon, 34141, South Korea. (e-mail: seunglee@kaist.ac.kr)}
\thanks{Iljoo Jeong is with the Department of Mechanical Engineering, Pohang University of Science and Technology (POSTECH), Pohang, 37673, South Korea. (e-mail: iljjeong@postech.ac.kr)}
}

\markboth{Journal of \LaTeX\ Class Files,~Vol.~xx, No.~x, xx~2024}%
{Shell \MakeLowercase{\textit{et al.}}: A Sample Article Using IEEEtran.cls for IEEE Journals}


\maketitle

\begin{abstract}
Accurate estimation of indoor space geometries is vital for constructing precise digital twins, whose broad industrial applications include navigation in unfamiliar environments and efficient evacuation planning, particularly in low-light conditions. This study introduces EchoScan, a deep neural network model that utilizes acoustic echoes to perform room geometry inference. 
Conventional sound-based techniques rely on estimating geometry-related room parameters such as wall position and room size, thereby limiting the diversity of inferable room geometries. Contrarily, EchoScan overcomes this limitation by directly inferring room floorplan maps and height maps, thereby enabling it to handle rooms with complex shapes, including curved walls.
\rev{The segmentation task for predicting floorplan and height maps enables the model to leverage both low- and high-order reflections. The use of high-order reflections further allows EchoScan}
to infer complex room shapes when some walls of the room are unobservable from the position of an audio device.
Herein, EchoScan was trained and evaluated using RIRs synthesized from complex environments, including the Manhattan and Atlanta layouts, employing a practical audio device configuration compatible with commercial, off-the-shelf devices.
\end{abstract}


\section{Introduction}

\IEEEPARstart{D}{gitial} twins have expedited innovative industrial applications spanning diverse sectors, such as navigating unfamiliar terrains or planning efficient evacuation blueprints \cite{liu_review_of_digital_twin}.
Room geometry is crucial information for rendering realistic audio in virtual reality (VR) and augmented reality (AR) environments, as well as for other sound-related applications such as source separation and sound field reconstruction. For example, in AR applications, the congruence of synthesized and real room impulse responses (RIRs) is essential for delivering an immersive audio experience \cite{intro_immersive_audio}, and room geometry information can help render realistic early reflections closely related to spatial audio perception. In sound source separation or enhancement tasks, knowledge of wall positions and corresponding image source locations has been reported to improve interference suppression performance greatly \cite{intro_RakingTheCocktailParty}. In sound field reconstruction problems, reconstructed sound fields by loudspeakers are distorted by room reflections \rev{\cite{Miyoshi1988Inv, elliott1989multiple, Santillan2001}}, and room geometry can provide important clues to suppress these reflections.     
Due to its importance in many tasks, significant research on room geometry inference (RGI) has been conducted in various ways using audio and vision sensor data.

\rev{In computer vision research, the inference of 3D room layouts from vision data has been approached in many ways.}  
Estimating the room layout from indoor RGB images has been tackled by identifying corners or boundaries of the floor and ceiling from indoor RGB images, which is crucial for a comprehensive understanding of a 3D scene. Panoramic images, offering a full 360° field of view and rich contextual information about a room, have demonstrated significant effectiveness in geometry estimation \cite{yang_dulanet,sun_horizonnet,pintore_atlantanet}. However, achieving accurate room geometry is challenging, particularly when visual data are limited or inaccessible. Cases with insufficient visual cues, such as disaster sites or power outages, highlight the inherent limitations of vision-based approaches. Even with visible light, estimating occluded geometries is fundamentally challenging for vision-based approaches. 

\begin{figure}[t]
    \centering
    \includegraphics[width=\columnwidth]{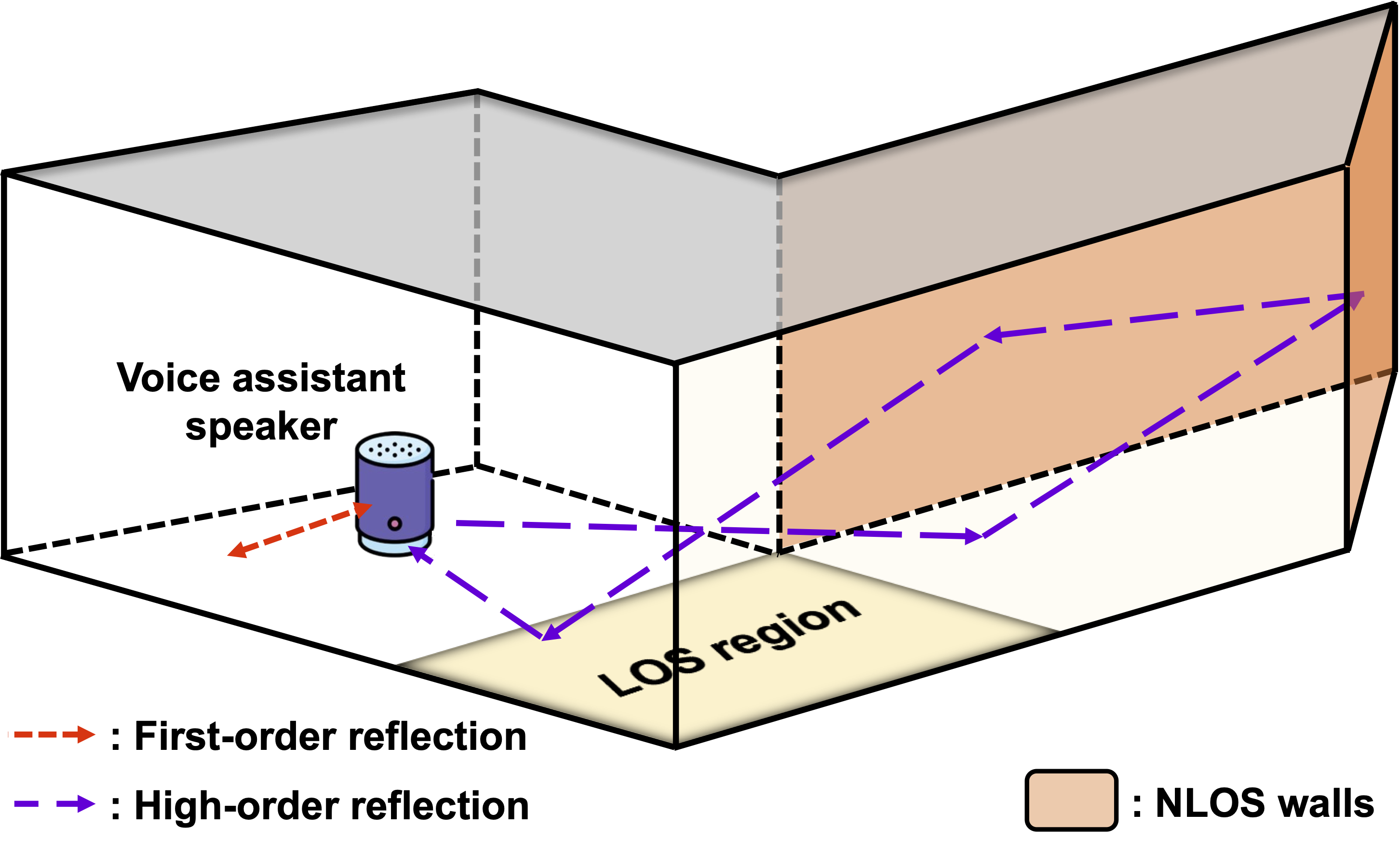}
    \caption{Conceptual illustration of the RGI task using an audio device positioned in the NLOS region.}
    \label{fig:overview}
\end{figure}

Acoustic echoes encapsulate essential information on geometrical characteristics of the room \cite{RIRcomponents}. When sound emitted from an audio device interacts with room boundaries, interactions such as specular and diffuse reflection, diffraction, and scattering are captured as RIRs. The representative features of RIRs for RGI are the time-of-arrival (TOA), which represents the duration required for sound to travel from a source to a receiver, and the direction-of-arrival (DOA), which indicates the impinging direction of the reflected waves.

For RGI, researchers have notably focused on TOAs of first-order reflections, which provide accurate distances to walls \cite{antonacci_inference_of_room_geometry, remaggi_a_3d_model_for_room_boundary_estimation, naseri_indoor_mapping, dokmanic_can_one_hear, dokmanic_acoustic_echoes_reveal, baba_3d_room_geometry_inference, remaggi_acoustic_reflector_localization_novel_image_source_reversion, rajapaksha_geometrical_room_geometry_estimation, park_iterative_echo_labeling, lovedee_three_dimensional_reflector_localisation}.
For example, ellipses can be formed by employing the collected TOAs of first-order reflections \cite{antonacci_inference_of_room_geometry, remaggi_a_3d_model_for_room_boundary_estimation} such that the two focal points of an ellipse correspond to the positions of a sound source and microphone. The boundary of the room can then be represented by a common tangential line across multiple ellipses. Remaggi \textit{et al.} \cite{remaggi_acoustic_reflector_localization_novel_image_source_reversion} compared several reflector localization techniques and showed that a direct localization model using the ellipsoid tangent sample consensus (ETSAC) performed better than other models. 
Dokmanic \textit{et al.} \cite{dokmanic_acoustic_echoes_reveal} presented an RGI technique based on the properties of the Euclidean distance matrix (EDM): a matrix of inter-microphone distances. They augmented an EDM for each image source based on pairwise distances to the microphones and conducted a rank test on the augmented matrices to obtain accurate echo combinations.
Lovedee-Turner and Murphy \cite{lovedee_three_dimensional_reflector_localisation} proposed an RGI method to overcome the convex-shape assumption required in most previous methods. They listed candidate walls from TOA-DOA pairs and filtered out impossible candidates through post-validation processes: path validation, line-of-sight (LOS) boundary validation, and closed geometry validation. Although this method can handle non-convex room geometries, it requires first-order reflections from every wall to be observable. 
Therefore, the microphone should be placed in the LOS region, where direct lines between the microphone and all the walls can be established, and the source should be relocated to multiple positions to obtain first-order reflections from every wall.

Several data-driven methods \cite{yu_room_acoustical_parameter_estimation, poschadel_room_geometry_estimation, gao_room_geometry_blind_inference, tuna_datadriven, yeon_3droom, yeon_rginet, luo_neural_acoustic_fields, purushwalkam_floorplan_reconstruction} have been proposed to overcome the limitations of conventional model-based methods that rely on low-order reflections. Deep neural networks (DNN) have been used to analyze the complex relationship between low- and high-order reflections.
Yu and Kleijn \cite{yu_room_acoustical_parameter_estimation} used a convolutional neural network (CNN) to analyze the relationship between the RIR and room acoustic parameters and estimated the size and absorption coefficient of quadrilateral rooms. Poschadel \textit{et al.} \cite{poschadel_room_geometry_estimation} employed a convolutional recurrent neural network (CRNN) to determine the lengths, widths, and heights of quadrilateral rooms using simulated RIRs. Tuna \textit{et al.} \cite{tuna_datadriven} also utilized a CRNN architecture to infer the 2D Cartesian coordinates of a real microphone and four imaginary microphones formed by sidewalls. This data-driven method exhibited RGI performance comparable to that of the model-based technique \cite{tuna_modeldriven} when tested with unseen measured RIRs.
Despite their outstanding RGI performance, these networks can only handle quadrilateral rooms because of their immutable \rev{number of output parameters}. Therefore, in our previous studies \cite{yeon_3droom, yeon_rginet}, we attempted to estimate various room geometries without considering the number of walls. This was possible by implementing an additional subnetwork that determines the confidence of the estimated wall parameters. However, the model cannot handle geometries with curved walls because it estimates the coefficients of the plane-wall equation. 
To address these challenges, we approach the RGI problem as a pixel segmentation task. This approach, inspired by vision-based methods \cite{yang_dulanet, pintore_atlantanet}, infers a 2D floorplan map and 1D height map sections of complex-shaped rooms, enabling the inference of room geometries with curved and non-line-of-sight (NLOS) walls.

The proposed EchoScan delivers three key contributions:
\begin{enumerate}[1)]
\item EchoScan handles more general and complex-shaped rooms, including those with curved and NLOS walls. RGI of \rev{geometrically complex} rooms is accomplished by a single compact audio device with an omnidirectional loudspeaker and circular microphone array. 

\item \rev{EchoScan aggregates echo-related latent features through a multi-aggregation (MA) module, which enables the model to compress latent features with multiple compression parameters.}

\item EchoScan fully utilizes high-order reflections, which is demonstrated through ablation studies and feature visualization analysis.
\end{enumerate}

\section{Problem Statement}\label{sec:problem}
Consider an indoor space or room surrounded by walls (Fig.\,\ref{fig:overview}), e.g., a meeting room or office room, in which an audio device comprising a loudspeaker and microphone array is placed at an arbitrary position. The sound emitted from the loudspeaker is reflected by walls and reaches the microphone array at different times. By analyzing these emitted and reflected sound waves, the acoustic fingerprint of a room, i.e., RIR, can be constructed. To secure practical accessibility for RIR measurement, this study considers an audio device that imitates an off-the-shelf voice assistant speaker with a single omnidirectional loudspeaker surrounded by microphones arranged in a circle with a fixed radius. We assume that the audio device can be placed within 70\% of length-width space of the given room to prevent the audio device from being positioned too close to the sidewalls and can be placed within the range of $[1, 1.5]$\,m from the floor. Details of assumptions and configurations of the audio device are addressed in Section\,\ref{sec:audio_device}.
 
The RGI problem can then be formulated as a geometric pixel segmentation task using the acquired RIRs. 
For the $i$-th 3D room ($i={1, \cdots, I}$), the input matrix $\bX_i \in \mathbb{R}^{M\times N}$ of the DNN model is given by $M$-channel RIRs with temporal length $N$ recorded by an audio device positioned at $\bp_i \in \mathbb{R}^3$ in the room. The output $\bY_i^\mathrm{3D}\in \mathbb{R}^{b\times b \times h}$ is a 3D tensor of segmented voxels defining a 3D room geometry centered at the position of the audio device. The output is sampled by $b$ pixels for length and width and $h$ pixels for height. Each segmented voxel contains binary values of 0 or 1, where 1 indicates the interior region of the room.
By assuming that the floor and ceiling are parallel to each other and perpendicular to the sidewalls, the 3D geometry $\bY_i^\mathrm{3D}$ can be represented as a combination of the 2D floorplan map $\bY_i^\mathrm{LW}\in \mathbb{R}^{b\times b}$ defined in the length-width space and 1D height map $\by_i^\mathrm{H}\in \mathbb{R}^h$ \cite{yang_dulanet, pintore_atlantanet}.
Using these definitions, the model can be trained to capture hidden information from $\bX_i$ to infer $\bY_i^\mathrm{LW}$ and $\by_i^\mathrm{H}$ despite the complexity of the floorplan maps and height maps.

Here, the floorplan maps $\bY_i^\mathrm{LW}$ and height maps $\by_i^\mathrm{H}$ are defined using local coordinates centered at the audio device position. This is necessary because RIRs do not contain information about the global coordinate system.
Accordingly, the audio device is always positioned at the center $(0,0)$ of the floorplan map, and the direction of the first microphone from the array center aligns with the negative length axis of the floorplan map. Also, even for the same room, different floorplan and height maps can be produced depending on the rotation angle and position of the audio device in the global coordinates.
 
In this study, we set the \rev{maximum allowable sizes of the floorplan and height maps} as $\pm$10.24\,m in length, $\pm$10.24\,m in width, and $\pm$5.12\,m in height from the audio device. The actual room that can be placed within these maps is smaller than the map size as detailed in Section\,\ref{sec:roomgeometry}.
This range was determined based on the typical sizes of indoor spaces where a voice assistant loudspeaker is commonly used, such as meeting rooms or living rooms. Specifically, the maps have pixel dimensions of 1024 pixels for each side of floorplan ($b=1024$) and 512 pixels for height ($h=512$), with an inter-pixel distance of 2\,cm. This inter-pixel distance was chosen to be smaller than the wall distance estimation errors reported in previous audio-based studies \cite{baba_3d_room_geometry_inference, lovedee_three_dimensional_reflector_localisation, yu_room_acoustical_parameter_estimation, tuna_datadriven}.

\begin{figure*}[t]
    \centering
    \includegraphics[width=\textwidth]{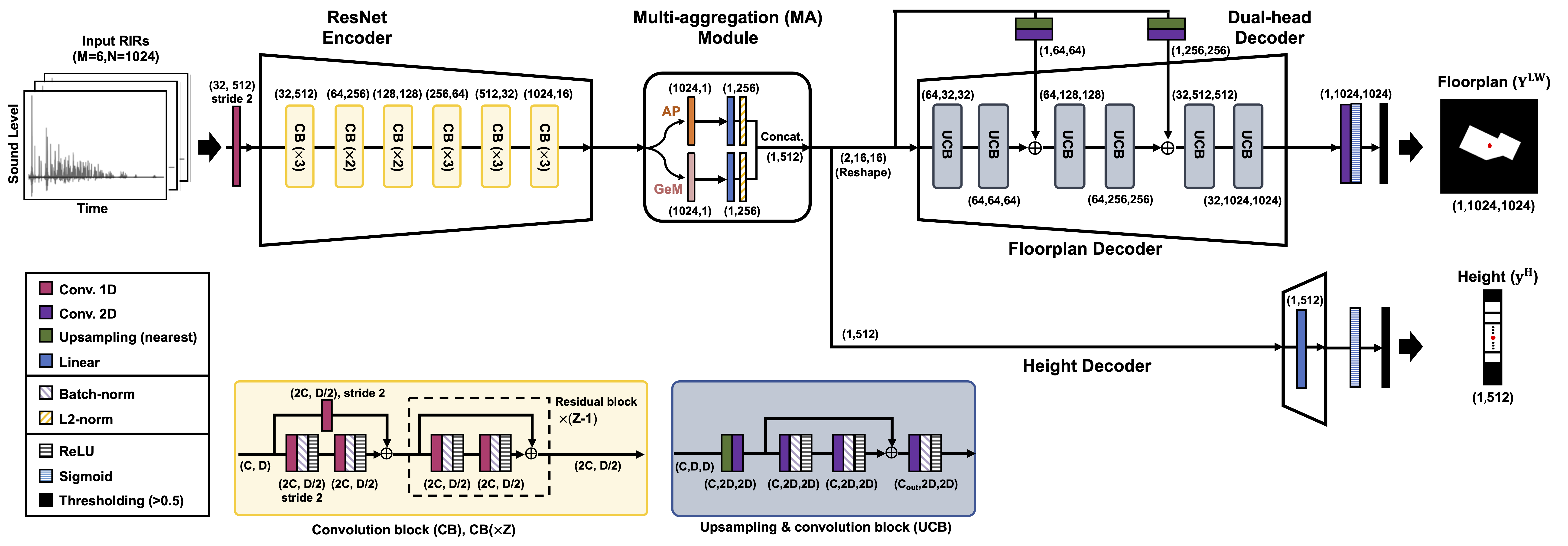}
    \caption{Encoder--decoder architecture of the proposed EchoScan. The encoder extracts latent features and the MA module aggregates them in time. The decoder generates two segmented images for the floorplan and height maps using its dual-head structure. The encoder consists of convolution blocks (CB) and the decoder comprises upsampling and convolution blocks (UCBs). The dimensions indicated with each encoder block or layer indicate its output dimensions (channel, time), while those for the decoder represent the output dimensions (channel, width, height). 
   The symbol $C$ denotes the channel dimension of the input, and $D$ is the time or space dimension for the 1D or 2D convolution block. For UCBs, the input is a 3D tensor with dimensions: (channels $C$, width $D$, height $D$), and the outputs are of size (channels $C_{out}$, width $2D$, height $2D$). Strides of convolution layers are 1 unless separately notified.
    } 
    \label{fig:model}
\end{figure*}

\section{Methodology}
\subsection{Encoder--Decoder Architecture}
This study proposes EchoScan, an encoder--decoder architecture for estimating the room geometry from RIRs, as shown in Fig.\,\ref{fig:model}. Among deep learning models, the encoder--decoder architecture is a well-established paradigm for cross-modal tasks \cite{hinton2006reducing,isola2017image}. This architecture is particularly effective for EchoScan, because EchoScan handles the cross-modal reconstruction task from audio to vision data. The encoder-decoder architecture can also encapsulate complex relationships between high-dimensional encoder inputs and decoder outputs into a reduced dimensionality of latent features. Accordingly, we can extract key geometry features in form of latent variables using this architecture.
The encoder extracts and compresses the spatio-temporal features $\bF$ from multichannel RIRs. This encoded latent features are aggregated by two distinct pooling operations in the MA module. Then, the aggregated features are separately fed into multi-head decoders to infer the 2D floorplan map and 1D height map.

As the encoder architecture, we employed ResNet \cite{he_resnet} to capture the relationships between reflections inherent in multichannel RIRs. 
The ResNet encoder has serial convolution blocks (CBs) with residual connection, which enables the model to learn high-level features without the gradient vanishing, commonly encountered in deep networks. This capability makes ResNet particularly advantageous for serving as the backbone in various sound-based research applications \cite{ma2021end,wang2023four}. For input RIRs with $M=6$ microphone channels and $N=1024$ samples in time, the first 1D convolution (1D Conv.) layer outputs $32$ channels and $512$ samples data using kernels of size 9. This kernel size was determined to cover 1\,ms of data, considering a sampling rate of 8\,kHz. Except for the first layer, all kernel sizes in the encoder were set to 5. The encoder comprises six CBs, each containing residual blocks with a residual connection every two layers. The number of residual blocks within each CB is indicated as $Z$ in `CB($\times Z$)' of Fig.\,\ref{fig:model}. In each CB, the number of channels ($C$) is doubled and the feature dimension ($D$) is halved except for the first CB. This transformation occurs in the first 1D Conv. layer and the first residual connection of each CB, with the stride set to 2. 
Except for these, the strides in the remaining 1D Conv. layers were set to 1, ensuring no change in channel size or dimension.
Each feature passing through the 1D Conv. layer undergoes batch normalization and activation via the rectified linear unit (ReLU). 
Features entering a residual block are summed with a residual connection after passing through the first two layers. Finally, the encoder outputs latent features $\bF=[\bf_1, \cdots, \bf_{C_{L}}]\T \in \mathbb{R}^{C_{L}\times D_{L}}$ with a channel dimension of $C_{L}=1024$ and a feature dimension of $D_{L}=16$.

Furthermore, the MA module was employed to aggregate features \rev{with different compression functions} \cite{jun_combination_of_multiple}. This module compresses the latent features $\bF$, through multiple global descriptors controlled by the compression parameter $\rho$, as defined by Equation~\eqref{eq:ma_module}.

\meq{
    {a_\rho(\bf_c)} = \Big( \frac{1}{\|\bf_c\|_0} \sum_{f \in \bf_c}f^{\rho} \Big)^{\frac{1}{\rho}}, \where c \in \{1, \cdots, C_{L}\}
}{eq:ma_module}
where $\|\mathbf{\cdot}\|_0$ represents the cardinality of a vector. When $\rho = 1$, the function equals {average-pooling (AP)}, compressing features {through global averaging}. In contrast, as $\rho \to \infty$, the function performs as a max-pooling, collecting only highly activated features. When $\rho =3$, the pooling function becomes generalized mean pooling (GeM), which moderately emphasizes strongly activated features and then aggregates{\cite{radenovic2018fine}. Both AP and GeM were utilized in the MA module to combine features aggregated with and without local emphasis}.
The latent output $\bF$ processed by AP and GeM results in two feature vectors of size \rev{$1024$}. 
Then, dimension reduction and normalization were performed using a linear layer and $\ell_2$-normalization, resulting in \rev{two groups of $256$ features}. These features were then concatenated to form an MA feature of size \rev{$512$}, which was used as input to the dual-head decoder.

The room geometry decoder infers the visual representation of a 3D room from the MA feature. We designed a dual-head decoder consisting of a floorplan decoder and a height decoder to generate a floorplan map (2D) and height map (1D) separately.
The floorplan decoder generates the predicted $b\times b$-pixel image $\hat{\bY}_i^\mathrm{LW}$.
The floorplan decoder includes a series of upsampling and convolution blocks (UCBs), each integrated with a residual connection to enhance feature propagation from the MA module. The first layer of each UCB has a nearest upsampling operator and a 2D convolution layer that effectively doubles the feature dimensions in both height and width. The input for the floorplan decoder is a reshaped MA feature of dimensions $2 \times 16 \times 16$, derived from aggregated features of size $512$.
In addition, \rev{projective} skip connections were integrated to reintroduce MA feature directly into the middle of the floorplan decoder. The skip connection is critical for directly propagating \rev{MA features to upper layers, similar to the DenseNet~\cite{Huang2017} architecture.} 
{To reconcile discrepancies in feature dimensions between the MA feature and the outputs of designated UCBs, specifically for the second and fourth UCBs, a nearest upsampling operation and a 2D convolution layer were utilized to align their sizes. Here, the upsampling factors for the second and fourth UCBs were 2 and 4 respectively.}

On the other hand, the height map decoder generates a $h$-pixel vector $\hat{\bY}_i^\mathrm{H}$ through a single linear layer.
{Since the ground truth (GT) floorplan and height maps contain binary values of 0 and 1, predicted values of the floorplan and height maps were \rev{mapped onto $[0, 1]$ by} a Sigmoid activation function located in the last layer of each floorplan and height decoder. During inference, the final binary floorplan and height maps were generated by applying a threshold of 0.5 to both outputs of Sigmoid activation function.}

\subsection{Loss Function}
\label{sec:pit}
The proposed model uses two types of loss function for training: mean squared error (MSE) and Dice loss. The MSE measures the average squared difference between the predicted values and the GT images. This guides the overall layout estimation of a given room.
The Dice loss function is useful for learning specific edge details in pixel segmentation tasks and measures the alignment between the predicted and GT layouts. The Dice loss function is given by Equation~\eqref{eq:dice_loss}.
\meq{
    L_{\operatorname{Dice}}^{\mathrm{LW}} = 
    \frac{1}{I} \sum_{i=1}^{I} 1 - \frac{2 (\hat{\by}_i^\mathrm{LW})\T \by_i^\mathrm{LW}}
    {\|\hat{\by}_i^\mathrm{LW} + \by_i^\mathrm{LW}\|_1},
}{eq:dice_loss}
where $\by^\mathrm{LW}$ is the vectorized form of the matrix $\bY^\mathrm{LW}$, and $\|\cdot\|_1$ is the 
{\(\ell_{1}\)-norm} of a vector.
The total loss function is given by the weighted sum of the MSE loss for the floorplan map, MSE loss for the height map, and Dice loss, as expressed in Eq.\,\eqref{eq:total_loss}.
\meq{
    L_\mathrm{total} = 
    L_{\operatorname{MSE}}^{\mathrm{LW}} +
    \alpha L_{\operatorname{Dice}}^{\mathrm{LW}} + 
    \beta L_{\operatorname{MSE}}^{\mathrm{H}}
}{eq:total_loss}
The weights $\alpha=0.3$ and $\beta=1$ were determined heuristically and showed good performances in all experiments. 

The circular microphone array used in this study does not distinguish ceiling reflections from floor reflections. To address this, we employed the permutation invariant training (PIT) technique \cite{pit}. It compares the estimated height vector with the original and flipped GT height vectors and updates the network with one that gives the lowest loss. After the inference is completed, the shorter side from the image center is considered the floor, because audio devices are usually positioned closer to the floor than the ceiling.

\section{Experiment Setup}
\subsection{{Audio Device Configuration}}
\label{sec:audio_device}
The audio device was configured using a circular microphone array of six omnidirectional microphones arranged on a ring with a loudspeaker placed at its center and a 5\,cm radius. The device was then randomly placed within 70\% of the length-width space of the room and a height range of $[1,~1.5]$\,m from the floor. The random positioning of the device is equivalent to a translated effect on the room configuration. Depending on the position of the audio device, the walls can be in either LOS or NLOS conditions. When straight-line connections can be made from the audio device to all walls without other walls obstructing, the room satisfies the LOS condition.

\subsection{{Acoustic Simulation}}
\label{sec:rir_simulation}
The raytracing engine of the Pyroomacoustics software \cite{pyroomacoustics} was employed to generate multichannel RIRs for general polyhedral rooms. Because the loudspeaker and microphones maintained consistent distances, the direct parts of the RIRs were omitted. RIRs were generated at an 8\,kHz sampling rate and included $N=1024$ samples in the time dimension. With this configuration, a single sample represents approximately 4.3\,cm of sound travel, and the total length of an RIR corresponds to 44\,m. Gaussian noise was added to emulate standard noise disturbances. The background noise was adjusted to ensure a signal-to-noise ratio (SNR) between $[10,~20]$~dB relative to the total energy of the RIR.

Wall absorption greatly affects the strength and dispersion of echoes. A set of typical absorption materials for floors, ceilings, and sidewalls defined in \cite{pyroomacoustics} were utilized and randomly assigned to each room. These materials include linoleum on concrete, carpet, and audience floor (wooden floor) for floors; gypsum boards, metal panels, and plasterboards for ceilings; and hard surfaces, rough concrete, rough lime washes, glass windows, and plasterboards for sidewalls.

\subsection{{Room Geometry Dataset}} \label{sec:roomgeometry}
We prepared two types of RIR datasets to extensively analyze and validate the proposed model: the RIR dataset of simple-shaped rooms (basic room dataset) and that of complex-shaped rooms ({Manhattan-Atlanta room dataset}). 
{ 
The basic room dataset consists of five specific types of rooms including convex and non-convex rooms, while the Manhattan-Atlanta room dataset includes these basic room shapes as well as a variety of complex room shapes, such as rooms with curved walls.}
The following sections describe the procedure for the room dataset construction.

\subsubsection{Basic Room Dataset}
\label{sec:Basic Room Dataset}
The basic room dataset includes RIRs simulated from five types of simple-shaped rooms: quadrilateral, pentagonal, hexagonal, L-type, and T-type. The dataset comprises \rev{a training dataset with 1,200,000 RIRs recorded at 200,000 locations and a test dataset with 6,000 RIRs recorded at 1,000 locations.}
Despite their simple shapes, all RIRs in this dataset were simulated in non-identical room configurations.
Room sizes were randomly chosen and an additional {distortion} step was introduced to diversify the aspect ratios. The rooms used for the test data belong to one of the five types; however, their \rev{vertex positions and sizes 
differ from those in the train dataset.}

{The room size parameter $\bs = [s_l,~s_w,~s_h]\T$, denoting length, width, and height, is a set of three numbers defining the size of a single room's floorplan and height maps. For a room with $K$-sided floorplan, we defined $K$ vertices $\bv_k\in \mathbb{R}^{2}$ using the size parameters $s_l$ and $s_w$ and construct a vertex matrix $\bV = [\bv_1, \cdots, \bv_K] \in \mathbb{R}^{2\times K}$ defining a 2D polygon. We then created a prototype room by extruding the 2D polygon in the height dimension by $s_h$. The room size parameters $s_l,~s_w,~s_h$ for each room layout were randomly selected from uniform distributions within the ranges $[2, 5]$, $[2, 5]$, and $[3, 5]$\,m, respectively. These ranges were determined by considering the small space where a voice assistant speaker is typically used.}
{With this configuration, the largest room possible is a quadrilateral room with dimensions of (10, 10, 5)\,m.}
For the {quadrilateral} rooms, the 2D vertex matrix was defined as 
\meq{
\bV^{Q} &= \begin{bmatrix}
-s_l &-s_l &s_l &s_l\\
-s_w &s_w  &s_w &-s_w
\end{bmatrix}.
}{}
For the pentagonal and hexagonal rooms, each vertex position $\bv_k$ of 2D vertex matrix was defined as 
\meq{
\bv_k &= \begin{bmatrix} s_l\cos \frac{2\pi k}{K},& s_w\sin \frac{2\pi k}{K} \end{bmatrix}\T.
}{}
For an L-type room, a {quadrilateral} room with lengths $s_l$ and $s_w$ was generated first, and then the cutout positions $\mu_l^L$ and $\mu_w^L$ were randomly determined within the ranges $[0, 0.5s_l]$ and $[0, 0.5s_w]$, respectively ($\bm{\mu}^L = [\mu_l^L, \mu_w^L]\T$). The 2D vertex matrix of the L-type rooms were defined as 
\meq{
\bV^{L} &= \begin{bmatrix} 
-s_l &-s_l &\mu_l^L &\mu_l^L &s_l   &s_l\\
-s_w &s_w  &s_w   &\mu_w^L &\mu_w^L &-s_w
\end{bmatrix}.
}{}
Similarly, for a T-type room, $\mu_l^{T1}$, $\mu_l^{T2}$, and $\mu_w^T$ were {chosen} randomly within the ranges $[-0.75s_l, -0.25s_l]$, $[0.25s_l, 0.75s_l]$, and $[-0.5s_w, 0]$ respectively ($\bm{\mu}^{T1} = [\mu_l^{T1}, \mu_w^T]\T$ and $\bm{\mu}^{T2} = [\mu_l^{T2}, \mu_w^T]\T$). The 2D vertex matrix of T-type rooms were defined as
\meq{
\bV^{T} = \begin{bmatrix} 
\mu_l^{T1} &\mu_l^{T1} &-s_l  &-s_l  &s_l &s_l   &\mu_l^{T2} &\mu_l^{T2}\\
-s_w        &\mu_w^T       &\mu_w^T &s_w   &s_w &\mu_w^T &\mu_w^T       &-s_w
\end{bmatrix}.
}{}
For every room, the audio device was randomly located between $[1, 1.5]$\,m from the floor and within 70\% space of a given 2D polygon ($0.7\bV$) defined by equally scaling down from every vertex.

{Further variations in room geometries were made by shifting the vertices of the generated prototype room shapes. Each vertex in the vertex matrix was randomly displaced by up to 0.5\,m along both the length and width axes. Consequently, each vertex was perturbed within a square of side length 0.5\,m centered on the original vertex position. This distortion step allows the model to accommodate various room shapes. Finally, the room shapes were rotated within the range $[0,~2\pi]$ in the length-width \rev{plane}. This rotation step enables the model to estimate room geometries aligned with the orientation of the audio device.} 

\begin{figure*}[t]
    \centering
    \includegraphics[width=\textwidth]{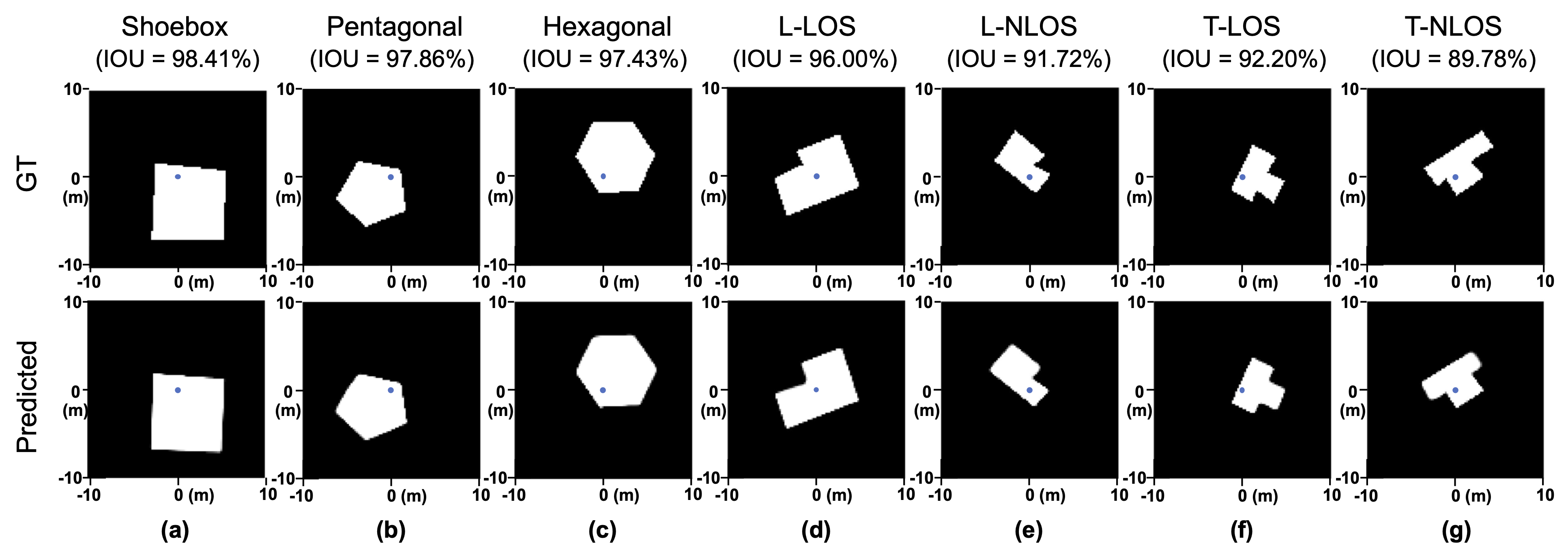}
    \caption{{Estimated floorplan and height maps for the basic room dataset containing five types of simple-shaped rooms: quadrilateral, pentagonal, hexagonal, L-type, and T-type. Two examples showing IOU performance close to the average IOU for their respective room types were selected and presented. The red dot indicates the position of the audio device, and the thick orange line displays the boundaries of the GT room. (a) Quadrilateral rooms, (b) Pentagonal rooms, (c) Hexagonal rooms, (d) L-LOS rooms, (e) L-NLOS rooms, (f) T-LOS rooms, and (g) T-NLOS rooms.}}
    \label{fig:results_standard}
\end{figure*}

A GT 2D floorplan was then generated by positioning a cross-sectional image of the generated room shape inside an image template of $b\times b$ pixels. As explained in Section\,\ref{sec:problem}, the generated room shape is centered at the origin of the local coordinates corresponding to the position of the audio device (Fig.\,\ref{fig:results_standard}).

\begin{figure}[t]
    \centering
    \includegraphics[width=\columnwidth]{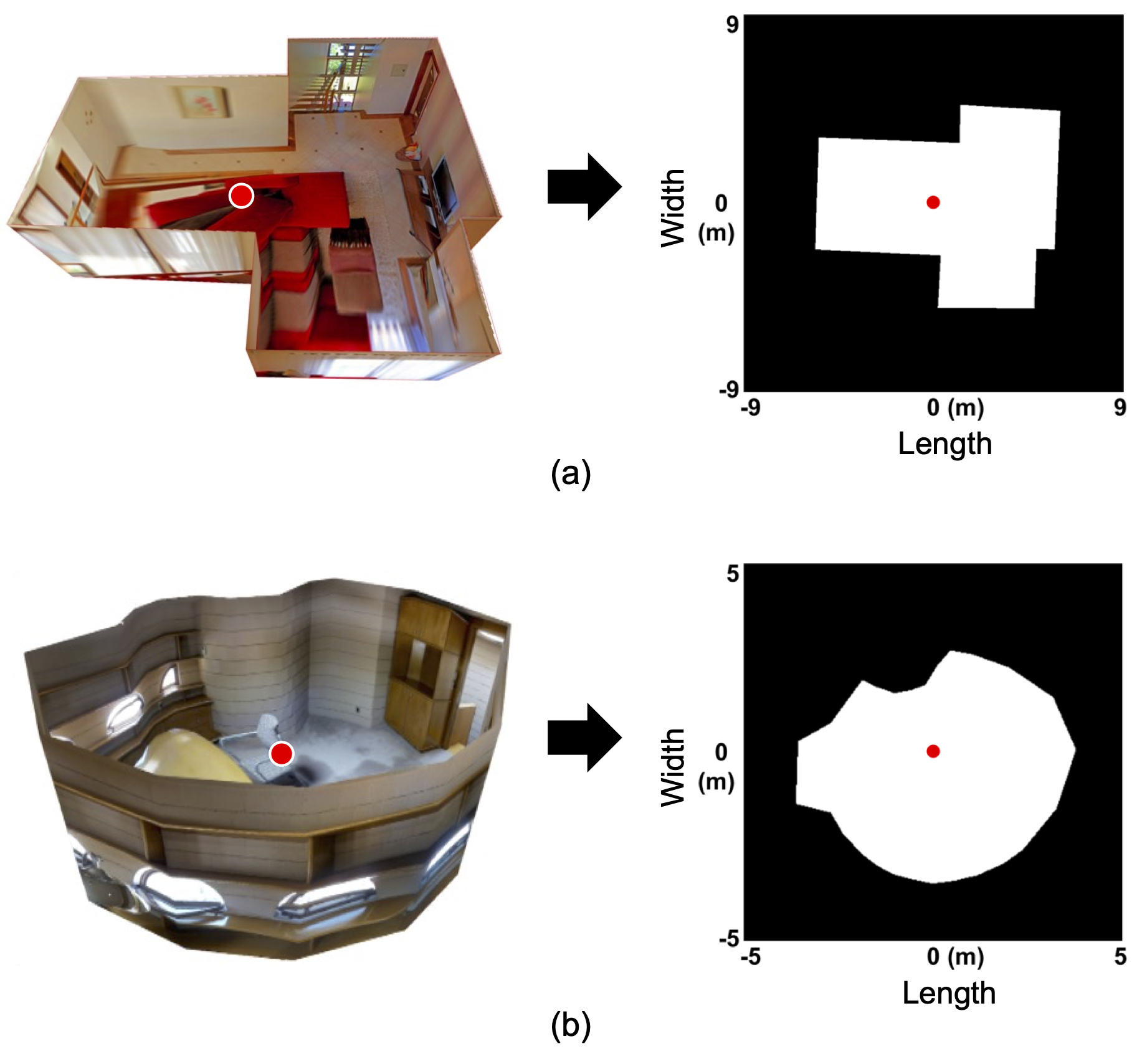}
    \caption{Examples of Manhattan and Atlanta layout rooms (left) and their floorplan maps (right). The red dots indicate the position of the audio device. Since EchoScan predicts the room geometry from the location of the audio device, the audio device is always at the center $(0,0)$ of the floorplan map. (a) Manhattan layout room containing only right-angled walls, and (b) Atlanta layout room including curved walls. Floorplan maps are magnified for better visibility.}
    \label{fig:example}
\end{figure}

\begin{figure}[t]
    \centering
    \includegraphics[width=\columnwidth]{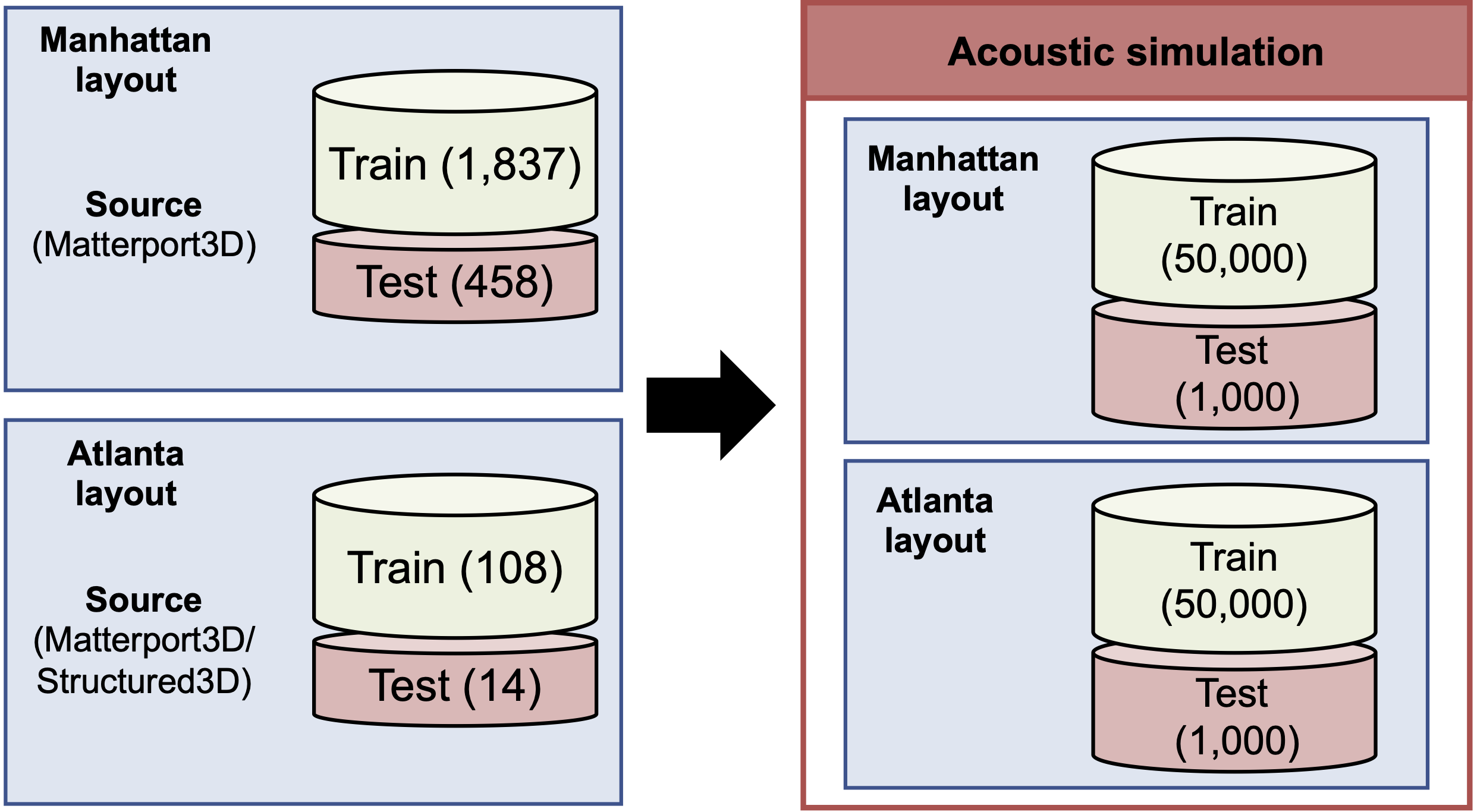}
    \caption{{Construction of the Manhattan-Atlanta room dataset. The room geometry dataset for both layouts follows the one used for the AtlantaNet \cite{pintore_atlantanet}. For each Manhattan and Atlanta layout, we simulated 50,000 RIRs for training and 1,000 RIRs for testing.}}
    \label{fig:Non-standardrooms}
\end{figure}

\subsubsection{{Manhattan-Atlanta} Room Dataset}
\label{sec:Manhattan-Atlanta Room Dataset}

To validate the performance of the proposed model on more realistic room geometries, we employed a publicly available room layout dataset with Manhattan-Atlanta room layouts. The Manhattan-Atlanta room dataset includes two types of layouts popularly utilized in vision-based approaches: the Manhattan layout (Fig.\,\ref{fig:example}(a)) \cite{coughlan_manhattan_world} and the Atlanta layout (Fig.\,\ref{fig:example}(b)) \cite{pintore_atlantanet}. {Both layouts presuppose that the sidewalls are orthogonal to the floors and ceilings. While Manhattan layouts necessitate that sidewalls also intersect at right angles, Atlanta layouts do not require such a constraint, thus enabling the inclusion of more complex and generic room shapes.}
{For Manhattan layouts, we selected rooms from the Matterport3D dataset, as described in \cite{zou20193d}. The room dataset for the Manhattan layout includes 1,837 rooms for training and 458 for testing (Fig.\ref{fig:Non-standardrooms}). For the Atlanta layout, we adopted the room dataset annotated in the AtlantaNet study \cite{pintore_atlantanet}, comprising 108 rooms for training and 14 for testing (Fig.\ref{fig:Non-standardrooms}). }

{To augment the Manhattan-Atlanta room dataset, we implemented a series of modifications to the existing room configurations. For the train dataset, we augmented the room sizes by applying two separate scaling factors for the floorplan and height maps, each randomly selected within a range of $[0.5, 2]$. The rotation and translation steps were also applied, as detailed in Section\,\ref{sec:Basic Room Dataset}. Rooms exceeding the \rev{maximum allowable} size of the floorplan and height maps after scaling and translation were removed from the Manhattan-Atlanta room dataset. For the test dataset, we only applied rotation and translation steps without room size augmentation. In both the train and test datasets, the audio device positions were set as the local coordinate center. Following this process, each dataset was augmented to include 50,000 train data and 1,000 test data as illustrated in Fig.\,\ref{fig:Non-standardrooms}. The resulting train datasets with 100,000 RIRs were used for fine-tuning the pre-trained model trained using the basic room dataset.}

\begin{table*}[t]
\centering
\caption{RGI Performance of Proposed and Ablation Models With Different {Temporal} Aggregation Functions}
\label{tab:rgi_sp_gem}
\resizebox{\textwidth}{!}
{
\large
\begin{tabular}{lrlccccccc|ccc}
\hline
\multicolumn{2}{c}{\multirow{2}{*}{Evaluation Metric}} &
  \multicolumn{1}{c}{\multirow{2}{*}{\begin{tabular}[c]{@{}c@{}}Model\end{tabular}}} &
  \multicolumn{3}{c}{Convex} &
  \multicolumn{4}{c|}{Non-convex} &
  \multicolumn{3}{c}{Average} \\
\multicolumn{2}{c}{} &
  \multicolumn{1}{c}{} &
  Quadrilateral &
  Pentagonal &
  Hexagonal &
  L-LOS &
  L-NLOS &
  T-LOS &
  T-NLOS &
  Convex &
  Non-convex &
  All \\ \hline
\multirow{3}{*}{IOU (\%)} &
  \multirow{3}{*}{$\uparrow$} &
  {AP}+GeM &
  {\textbf{98.50}} &
  {\textbf{97.63}} &
  {\textbf{97.30}} &
  {\textbf{96.03}} &
  {\textbf{95.34}} &
  {\textbf{95.13}} &
  {\textbf{92.15}} &
  {\textbf{97.81}} &
  {\textbf{94.66}} &
  {\textbf{96.01}} \\
 &
   &
  {AP} &
  {96.96} &
  {95.47} &
  {95.10} &
  {93.78} &
  {92.24} &
  {91.44} &
  {88.39} &
  {95.84} &
  {91.46} &
  {93.34} \\
 &
   &
  GeM &
  {97.70} &
  {96.64} &
  {96.20} &
  {94.38} &
  {93.68} &
  {92.72} &
  {89.57} &
  {96.85} &
  {92.59} &
  {94.41} \\ \hline
\multirow{3}{*}{$\text{MSE}_{LW}~(\times 10^{-3})$} &
  \multirow{3}{*}{$\downarrow$} &
  {AP}+GeM &
  {\textbf{2.9}} &
  {\textbf{3.2}} &
  {\textbf{3.5}} &
  {\textbf{6.7}} &
  {\textbf{7.6}} &
  {\textbf{8.3}} &
  {\textbf{13.0}} &
  {\textbf{3.1}} &
  {\textbf{8.9}} &
  {\textbf{6.4}} \\
 &
   &
  {AP} &
  {5.9} &
  {5.4} &
  {6.4} &
  {10.2} &
  {11.8} &
  {14.7} &
  {19.4} &
  {5.9} &
  {14.03} &
  {10.54} \\
 &
   &
  GeM &
  {4.2} &
  {3.9} &
  {4.8} &
  {9.4} &
  {9.7} &
  {12.4} &
  {17.2} &
  {4.3} &
  {12.18} &
  {8.8} \\ \hline
\multirow{3}{*}{$\text{MSE}_{H}~(\times 10^{-3})$} &
  \multirow{3}{*}{$\downarrow$} &
  {AP}+GeM &
  {\textbf{0.9}} &
  {\textbf{0.8}} &
  {\textbf{0.8}} &
  {\textbf{0.8}} &
  {\textbf{0.8}} &
  {\textbf{0.8}} &
  {\textbf{1.0}} &
  {\textbf{0.8}} &
  {\textbf{0.9}} &
  {\textbf{0.8}} \\
 &
   &
  {AP} &
  {1.5} &
  {1.5} &
  {1.4} &
  {1.5} &
  {1.5} &
  {1.4} &
  {1.6} &
  {1.5} &
  {1.5} &
  {1.5} \\
 &
   &
  GeM &
  {1.8} &
  {1.6} &
  {1.8} &
  {2.0} &
  {1.5} &
  {1.9} &
  {1.9} &
  {1.7} &
  {1.8} &
  {1.8} \\ \hline
\end{tabular}%
}
\end{table*}

\subsection{Training Configuration}
Four NVIDIA GeForce RTX {A6000} GPUs were used to train the model with a batch size 32. \rev{The EchoScan model was first trained on the basic room dataset for 300 epochs (approximately 468,000 iterations) and then fine-tuned on the Manhattan-Atlanta room dataset for 150 epochs (approximately 117,000 iterations).} The learning rate was varied using the cosine-annealing warmup restart scheduler \cite{scheduler} with an initial learning rate of {$10^{-3}$} to a minimum of {$10^{-5}$}. The Adam optimizer \cite{adam} was used for backpropagation.
During training, variable-length time masking was applied to the input RIRs \cite{specaugment} to enhance the robustness of the model. Three masks were applied to random temporal locations of RIRs, and the lengths of the masks were randomly chosen within $[0,~100]$ samples.

\subsection{Evaluation Metrics}
The performance of the proposed model was verified using two performance evaluation metrics. First, MSE was used to evaluate both floorplan and height values. Second, intersection over union (IOU) was adopted as a similarity measure at the pixel or voxel level. IOU for the room geometry can be defined as
\meq{
    \operatorname{IOU} = 
    \frac{1}{I} \sum_{i=1}^{I}
    \frac{(\hat{\by}_i^{\mathrm{3D}})\T \by_i^{\mathrm{3D}}}
    {\|\hat{\by}_i^{\mathrm{3D}} + \by_i^{\mathrm{3D}}\|_1 - (\hat{\by}_i^{\mathrm{3D}})\T \by_i^{\mathrm{3D}}},
}{eq:metric_iou}
where $\by^{\mathrm{3D}}$ is the vectorized form of the matrix $\bY^{\mathrm{3D}}$. {The IOU is given by the ratio of an overlapping region to the combined region, calculated from pixels of two segmented images. This metric provides a direct measure of \rev{geometrical} similarity between the estimated and GT room shapes. Since the significance of the mismatch between the predicted and GT geometries should vary depending on the absolute size of the room, the IOU, defined as a ratio, effectively conveys the significance of the mismatch, accounting for the absolute room size.
}

\section{Experimental Results}
\subsection{Ablation Studies on the MA Module}
To demonstrate the effectiveness of the proposed model and its MA modules, we conducted ablation studies with different combinations of aggregation functions: {MA module (AP+GeM) and single-aggregation modules (AP only or GeM only).} {The comparison results in Table~\ref{tab:rgi_sp_gem} indicate that when using single-aggregation, AP and GeM exhibit almost similar levels of accuracy, with only approximately a 1\% difference in IOU. However, when using the MA module with both aggregations, IOU is improved by up to 3.7\% compared to AP and GEM for the T-LOS and T-NLOS rooms. For convex rooms, the performance improvement is rather subtle (1--2\%) compared to non-convex rooms (2--4\%). However, the consistent improvement across all room shapes demonstrates that aggregating both the locally emphasized features (GeM) and globally averaged features (AP) in RIRs is beneficial for identifying various room shapes.}

\begin{figure*}[t]
    \centering
    \includegraphics[width=\textwidth]{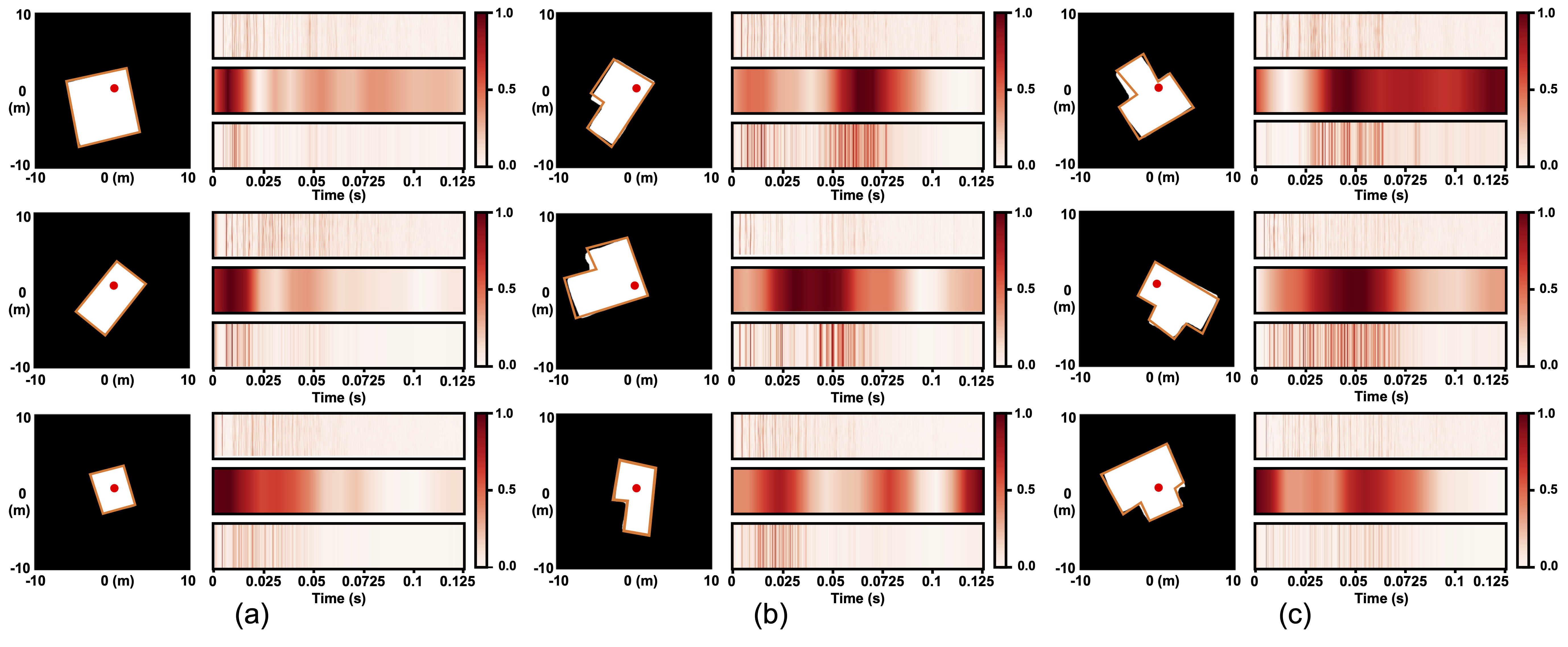}
    \caption{{Visualization of temporal activation using Grad-CAM. (a), (b), and (c) display the Grad-CAM results for three different rooms selected from quadrilateral, L-type, and T-type room types, respectively. In each case, the left image shows the estimated floorplan map with the thick orange line and red dot indicating the GT room shape and audio device position, respectively. 
    The three graphs on the right show, in order: the input 6-channel RIRs, the Grad-CAM activation map, and the highlighted RIRs. The highlighted RIRs are derived by taking the maximum value across the channel dimension from the product of the input RIRs and the Grad-CAM activation. Each graph is normalized to its maximum value.}}
    \label{fig:gradcam}
\end{figure*}

\subsection{Performance Analysis for Basic Room Dataset}
The performance of the proposed model in estimating the geometry of 3D indoor space is presented in Fig.\,\ref{fig:results_standard} and Table\,\ref{tab:rgi_sp_gem}. The overall result shows an IOU of more than 90\% for all types of basic rooms and a negligible MSE for height estimation. This indicates that the height estimation task is sufficiently simple to accomplish using a decoder with a single linear layer.
First-order reflections present distinct peaks in RIRs; therefore, encoding features related to first-order reflections requires minimal effort. Thus, the {quadrilateral} rooms showed the best performance for both MSE (LW) and IOU, although it tended to decrease in non-convex room shapes, for which some first-order reflections were missing in the measured RIRs.
In addition, the performance in the LOS and NLOS cases of T-type rooms was notably different, even for rooms of the same type.

Table\,\ref{tab:rgi_sp_gem} and Fig.\,\ref{fig:results_standard} show RGI results for basic rooms. Samples with a similar IOU to the mean IOU were selected.
For the convex rooms illustrated in Figs.\,\ref{fig:results_standard}(a)--(c), the model could accurately predict the room shapes. For the non-convex rooms shown in Figs.\,\ref{fig:results_standard}(d)--(g), although the estimations were less accurate than those for convex rooms, the accuracy was still high.
In particular, for the non-convex NLOS rooms shown in Figs.\,\ref{fig:results_standard}(e) and (g), rooms were overestimated or underestimated because some walls are invisible from the position of the device. 
However, the overall shapes, including invisible walls, can still be estimated, which was impossible with vision- or previous sound-based approaches. These results highlight the significance of using high-order reflection information based on sound propagation characteristics to predict NLOS-type rooms.

\subsection{Role of High-Order Reflections}
To assess whether our model considers higher-order reflections, we conducted two distinct investigations: model interpretation and data analysis.
For model interpretation, gradient-weighted class activation mapping (Grad-CAM) \cite{gradcam} was utilized to highlight the temporal areas in the input RIRs that affect the model predictions.
{Fig.\,\ref{fig:gradcam}(a) shows three different examples of a simple quadrilateral room. In these cases satisfying the LOS condition, most activations occurred in the early temporal region where the first-order reflections were recorded. These results are in line with conventional sound-based approaches that only exploit visible early reflections.} 
{Figs.\,\ref{fig:gradcam}(b) and (c) show three examples for L-type and T-type rooms, respectively. For these more complex rooms, activation becomes strong in the later temporal regions where high-order reflections dominate, compared to the cases of simple quadrilateral rooms.
The temporal interval from 0.05\,s to 0.1\,s corresponds to the sound travel distance of approximately 17\,m to 34\,m for high-order reflections. These later activations signify that the model actively utilizes high-order reflections to estimate complex room geometries.}

\begin{figure}[t]
    \centering
    \includegraphics[width=\columnwidth]{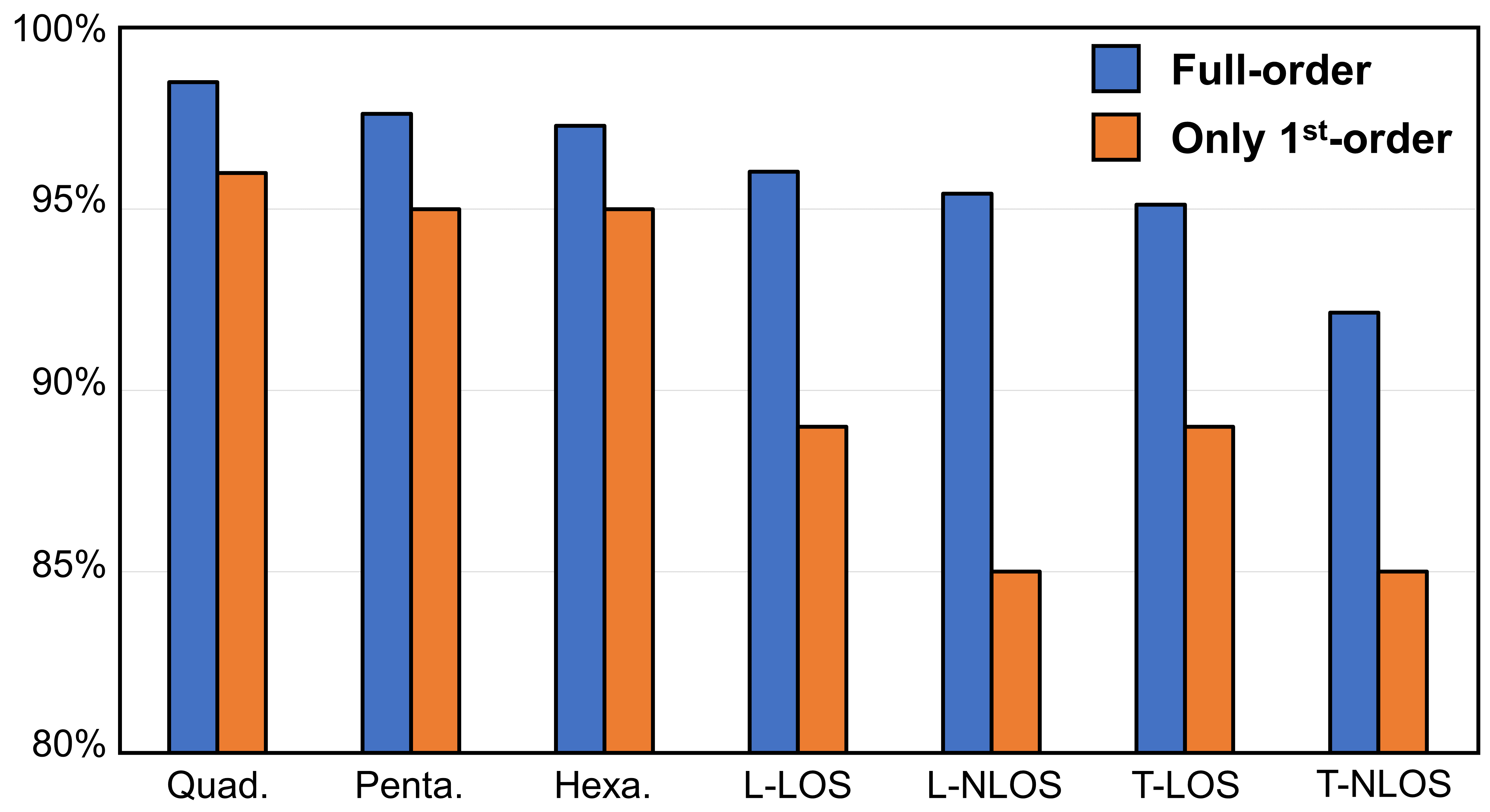}
    \caption{Quantitative performance (3D IOU) comparison between RIRs with full-order and only first-order reflections. The RGI performance significantly decreases when estimating L- and T-type rooms using RIRs that contain only first-order reflections. This indirectly suggests that EchoScan utilizes high-order reflections.}
    \label{fig:without_high_order_graph}
\end{figure}

\begin{figure}[t]
    \centering
    \includegraphics[width=0.8\columnwidth]{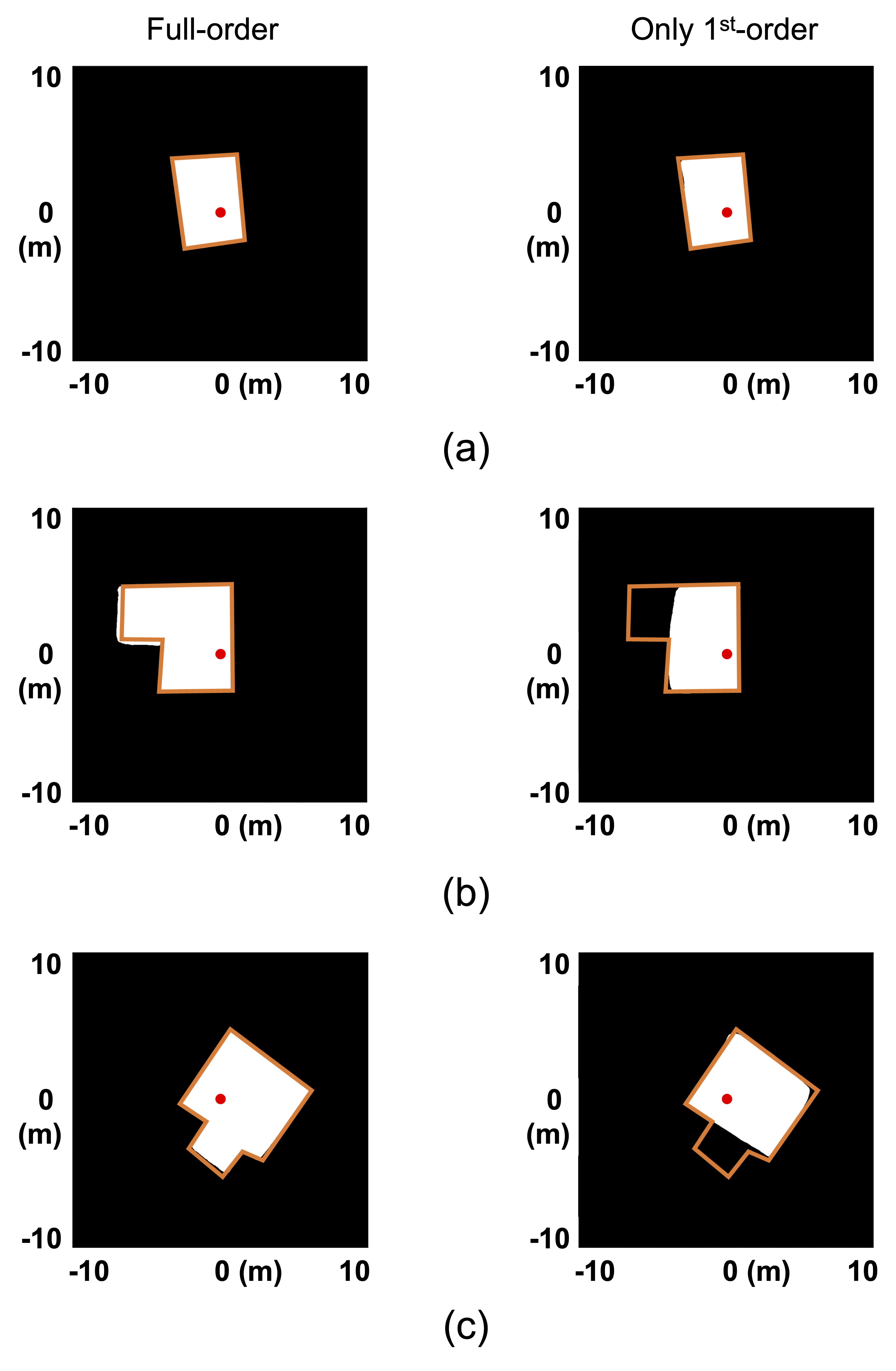}
    \caption{{Comparison of the predicted floorplan maps between RIRs with full-order (left) and only first-order (right) reflections. The thick orange line displays the boundaries of the GT room. (a) Quadrilateral room, (b) L-NLOS room, and (c) T-NLOS room. The L- and T-type rooms are estimated as quadrilateral rooms when the reflection order of RIRs is truncated to include only first-order reflections.}}
    \label{fig:without_high_order_floorplan}
\end{figure}

To further confirm the utilization of high-order reflections, we compared the performance of two distinct EchoScan models trained using RIRs with full- and first-order reflections. The results are shown in Figs.\,\ref{fig:without_high_order_graph} and \ref{fig:without_high_order_floorplan}, which reveal a notable disparity. When the {input to the} model was constrained to first-order reflections, its IOU score for non-convex rooms decreased considerably, particularly in NLOS scenarios, compared to the model trained with full-order reflections. Specifically, the model trained with limited reflections could accurately estimate a {quadrilateral} room, as shown in Fig.\,\ref{fig:without_high_order_floorplan}(a), {but} it struggles with more complex geometries such as a {non-convex L- and T-type rooms (Figs.\,\ref{fig:without_high_order_floorplan}(b) and (c))}, {often oversimplifying} a room in a basic {quadrilateral} shape. This result also indicates that EchoScan uses high-order reflections to infer complex room geometries.

\begin{figure*}[t]
    \centering
    \includegraphics[width=\textwidth]{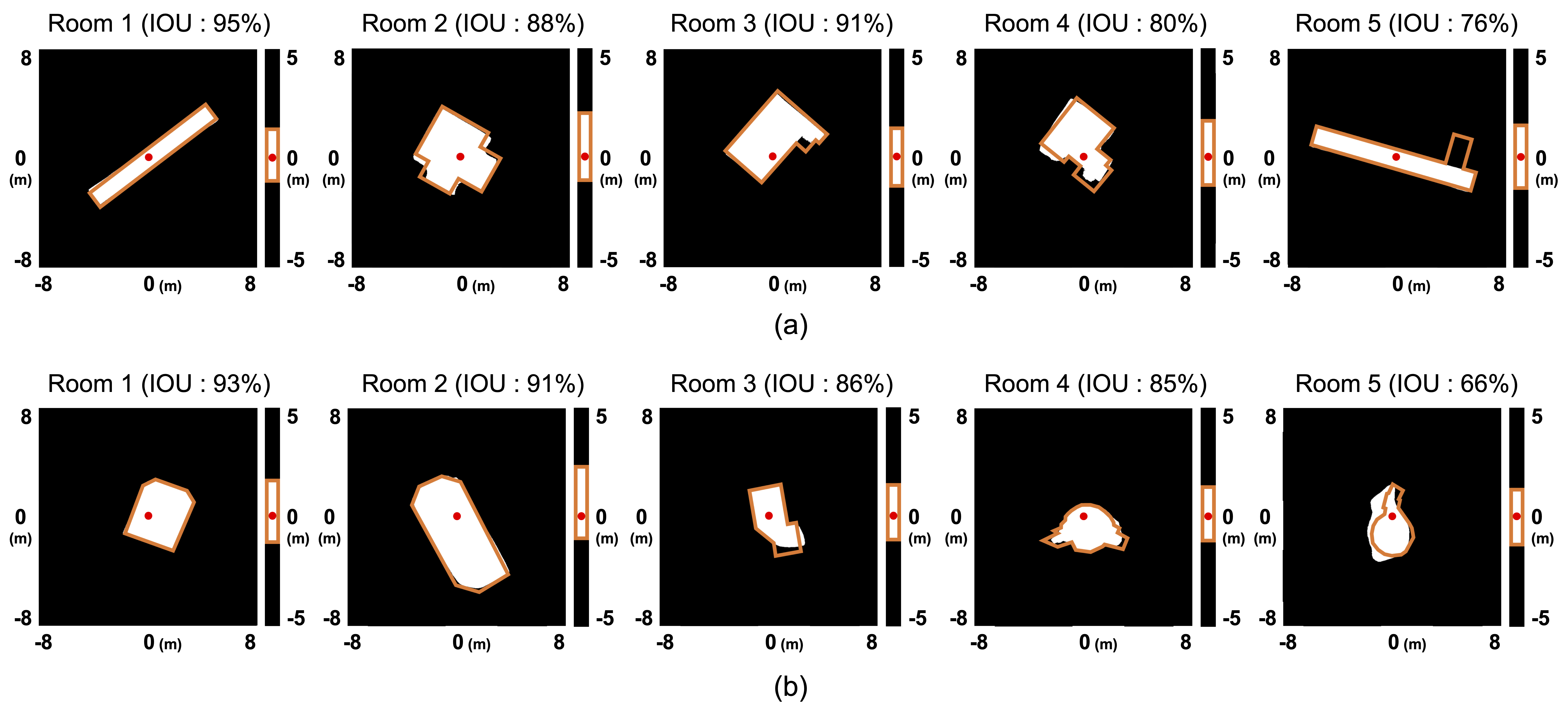}
    \caption{{RGI examples from the Manhattan-Atlanta room dataset. The red dot and thick orange line denote the position of the audio device and the boundaries of the GT room, respectively. (a) Manhattan layout rooms and (b) Atlanta layout rooms. Floorplan maps are uniformly zoomed into a range of [-8, 8] m for better visibility.}}
    \label{fig:good_example}
\end{figure*}

\subsection{Performance Analysis for {Manhattan-Atlanta Room Dataset}}
{The RGI performance for the Manhattan and Atlanta layouts is shown in Table~\ref{tab:rgi_vision}. Since no previous acoustic-based model can infer the complex geometry of these Manhattan-Atlanta rooms, we presented the results with the ones reported for the vision-based models \cite{yang_dulanet, pintore_atlantanet, sun_horizonnet}. 
The configuration of room layouts and train/validation/test dataset splits were identical to those used in \cite{pintore_atlantanet}. 
\rev{For DuLa-Net \cite{yang_dulanet} and HorizonNet \cite{sun_horizonnet}, we referenced the inference results reported in the AtlantaNet paper \cite{pintore_atlantanet}. For AtlantaNet, we used the version with the ResNet50 encoder, as specified in the original paper.} 

Like vision-based models, Echoscan shows the highest performance for simple geometries (Manhattan layout with four corners), and as the room geometries become more complex with more corners, the performance is reduced. EchoScan shows similar or even higher performance than vision-based models in most room types, and its parameter size and computational complexity presented in MAC (multiply-accumulate) operations are also comparable to those of vision-based models.
However, fundamental domain-wise differences in the input data should be considered. For example, the input data to the EchoScan model are multichannel RIRs simulated without furniture or indoor objects, whereas inputs to the vision-based models are 360$^{\circ}$ panorama images with various types of furniture (Additional RGI results for RIRs with indoor objects are described in Section\,\ref{sec:funiture}.). Therefore, the interpretation we can draw from these results is that the RGI accuracy of the proposed model can be comparable to or higher than those of vision-based models in these simplified conditions through the inference of NLOS walls using high-order reflections.}

{Fig.\,\ref{fig:good_example}(a) showcases the RGI results for five rooms with Manhattan layouts. The first room, Room 1, is a long, narrow corridor-shaped quadrilateral room, approximately 10\,m in length. 
For Rooms 2 to 4, EchoScan captures the primary room dimensions and overall layout with some errors around corners and bent sections. However, for Room 5 characterized by a long corridor ending in a right-angled T-shaped bend, EchoScan fails to infer the angled portion. The results for Rooms 2 to 5 indicate that complex right-angled structures or long acoustic propagation paths requiring extensive high-order reflections can be challenging for acoustic-based RGI.}

{Fig.\,\ref{fig:good_example}(b) presents the estimation results for five rooms with Atlanta layouts. Room 1 and Room 2 featuring convex Atlanta shapes are accurately estimated. In the case of Room 3, although not completely accurate, EchoScan captures the general shape including concave sections. Even for the significantly curved room (Room 4), the room shape is accurately estimated except for small areas near the edges. These results underscore EchoScan’s ability to predict complex Atlanta layouts with curved walls. The worst case among these examples is Room 5 consisting of completely curved walls on the lower side, \rev{near $(0,\,-2)$\,m in the 2D coordinates}, and a small chamber in the upper section, \rev{around $(0,\,+1.5)$\,m}. We can see the most significant discrepancy in the estimated room shape, which is attributable to scattered reflections from the curved walls arriving at similar times and mixed with the reflections from the upper section.}

\begin{table}[t]
\centering

\begin{threeparttable}
\caption{Performance Comparison (3D IOU, \%) of Proposed and Vision-based Methods on Real-world Room Dataset}
\label{tab:rgi_vision}
\begin{tabular}{lrccc|c}
\hline
\multicolumn{2}{l}{\begin{tabular}[c]{@{}l@{}}Room type\\ (\# of corners)\end{tabular}} &
  \begin{tabular}[c]{@{}c@{}}DulaNet\\ \cite{yang_dulanet,pintore_atlantanet}\end{tabular} &
  \begin{tabular}[c]{@{}c@{}}HorizonNet\\ \cite{sun_horizonnet,pintore_atlantanet}\end{tabular} &
  \begin{tabular}[c]{@{}c@{}}AtlantaNet\\ \cite{pintore_atlantanet}\end{tabular} &
  Ours \\ \hline
Manhattan       & (4)           & 77.0   & 81.9   & 82.6    & {95.1}   \\
Manhattan       & (6)           & 78.8   & 82.3   & 80.1    & {86.6}   \\
Manhattan       & (8)           & 71.0   & 71.8   & 71.2    & {81.4}   \\
Manhattan       & ($>$10)       & 63.3   & 68.3   & 73.9    & {72.6}   \\ \hline
Atlanta         & (6)           & --     & 74.5   & 84.3    & {88.3}   \\
Atlanta         & (8)           & --     & 65.0   & 78.4    & {85.3}   \\
Atlanta         & ($>$10)        & --     & 64.4   & 75.3    & {79.3}   \\ \hline
\multicolumn{2}{l}{{Param. Size}} & {25.6 M} & {81.6 M} & {100.2 M} & {44.1 M} \\
\multicolumn{2}{l}{{MACs}}        & {46.8 G} & {71.9 G} & {273.7 G} & {49.0 G} \\ \hline
\end{tabular}

\begin{tablenotes}
      \footnotesize
      \item {* This table is not for direct comparison across different models. Rather, we aim to demonstrate that the acoustic-based method (ours) can perform similarly to vision-based methods for RIRs simulated by raytracing in the absence of indoor objects. Here, we used the same output dimensions $(1024\times 1024)$ for the floorplan maps and utilized the same room geometry dataset as \cite{pintore_atlantanet} for training and testing to provide similar information to DNN models as much as possible. However, for the vision-based models, the input is the panorama image with furniture and indoor objects, whereas the input is clean multichannel RIR for the acoustic-based model. \rev{Furthermore, in vision-based approaches, the inter-pixel distance of the floorplan map varies depending on the height of the camera position relative to the height of the room.}}
    \end{tablenotes}

\end{threeparttable}

\end{table}

\begin{figure}[t]
    \centering
    \includegraphics[width=\columnwidth]{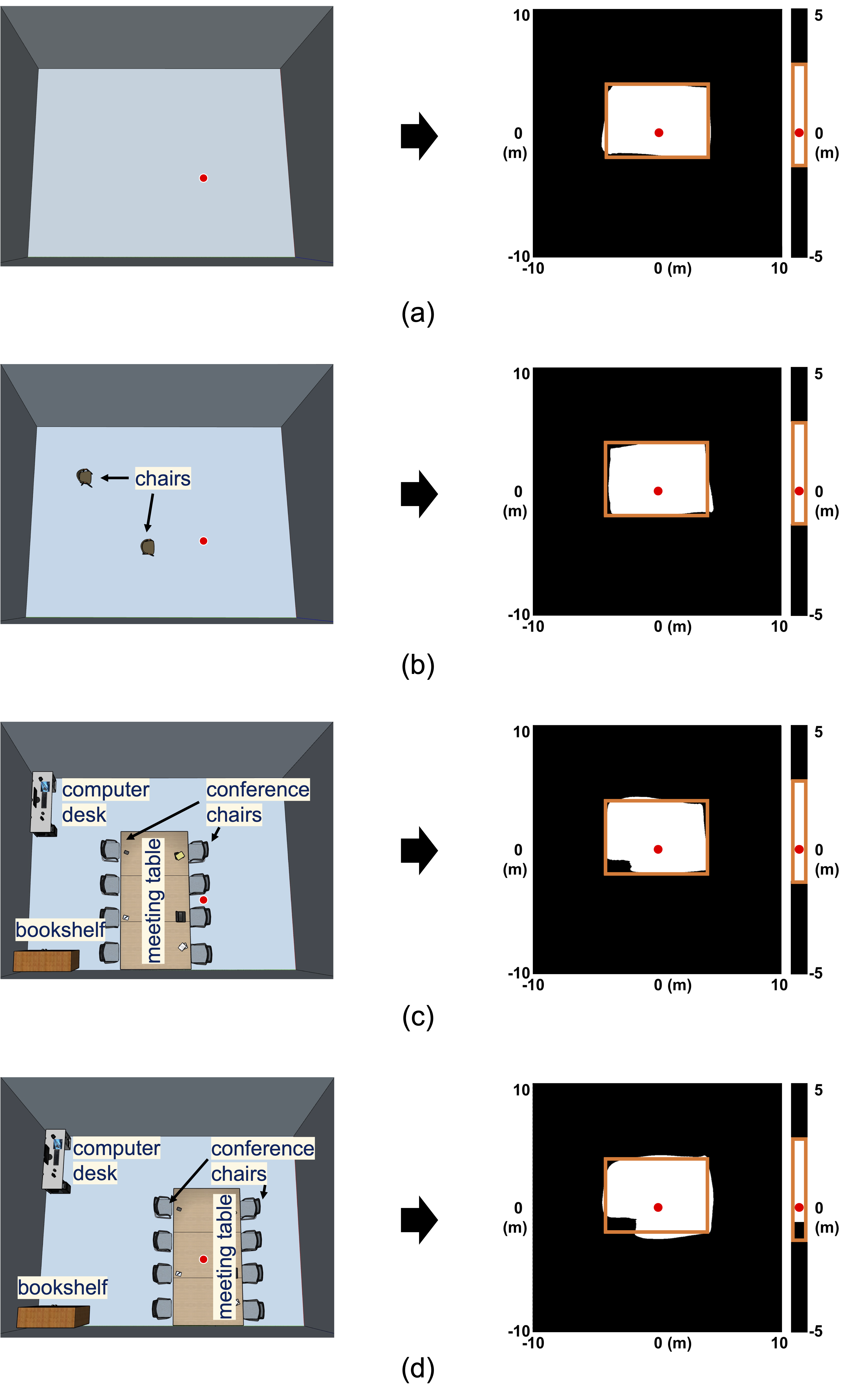}
    \caption{{A top view of the rooms with various furniture arrangements (left) and their corresponding floorplan maps and height maps (right). The thick orange line denotes the boundaries of the GT room. (a) A room without furniture, (b) a room with minimal furniture (two chairs), (c) a room with diverse types of furniture (a meeting table, conference chairs, a computer desk, and a bookshelf), and (d) a room with the same furniture as in (c) but with the meeting table positioned beneath the device.}}
    \label{fig:furniture}
\end{figure}

\subsection{{Experiments With Indoor Objects}}
\label{sec:funiture}
{
Indoor objects such as furniture scatter acoustic waves propagating in the room. To investigate the effects of indoor objects on the estimated geometry, experiments were carried out by varying types and arrangements of objects in the same room. Since Pyroomacoustics \cite{pyroomacoustics} does not support RIR simulation with scattering objects, the ODEON\texttrademark~room acoustics software \cite{odeon} was utilized to simulate RIRs. SketchUp\texttrademark~software was used to build the 3D model of the room with objects, as illustrated in Fig.\,\ref{fig:furniture}.
}

{
Four cases were considered in a quadrilateral room of size $(8, 6, 4)$\,m: the first case only including an empty room without indoor objects (Fig.\,\ref{fig:furniture}(a)), the second case with only small objects such as chairs (Fig.\,\ref{fig:furniture}(b)), the third case with small and large furniture like a big meeting table and a bookshelf (Fig.\,\ref{fig:furniture}(c)), and the fourth case with small and large furniture and the audio device positioned above the meeting table (Fig.\,\ref{fig:furniture}(d)). The audio device was placed at $(4.5, 2, 1.25)$\,m relative to the origin of the global coordinates set in the lower left corner of the room. The source-microphone configuration and preprocessing of the simulated RIRs follow those described in Sections\,\ref{sec:audio_device} and \ref{sec:rir_simulation}, respectively. The acoustic materials for the walls and furniture are selected in the global material library of ODEON\texttrademark~room acoustics software. The materials used in this experiment and their absorption coefficients across eight-octave bands (63\,Hz to 8\,kHz) are as follows: the painted concrete (ID 103) with absorption coefficients of $[0.1, 0.1, 0.1, 0.1, 0.1, 0.1, 0.1, 0.1]$ for the sidewalls; the 10\,mm soft carpet (ID 7007) with coefficients of $[0.1, 0.1, 0.1, 0.2, 0.3, 0.3, 0.4, 0.4]$ for the floor; the 27\,mm gypsum board (ID 4053) with coefficients of $[0.5, 0.5, 0.6, 0.6, 0.9, 0.9, 0.8, 0.8]$ for the ceiling; the chairs with cloth covers (ID 11006) with coefficients of $[0.4, 0.4, 0.6, 0.8, 0.9, 0.8, 0.7, 0.8]$ for the all types of chairs; and the 25\,mm wood panel (ID 3065) with coefficients of $[0.2, 0.2, 0.1, 0.1, 0.1, 0.1, 0.1, 0.1]$ for the meeting tables, computer desks, and bookshelves. The scattering coefficient for simulating diffuse reflections is 0.1 for all materials.
No additional fine-tuning steps were applied in these experiments to account for differences between simulation tools.
}

{
Figs.\,\ref{fig:furniture}(a) and (b) present the 3D models of rooms without and with two chairs, respectively, together with the predicted floorplan and height maps.
The results show that small objects do not seriously degrade the estimated room geometries. In the next experiment shown in Figs.\,\ref{fig:furniture}(c) and (d), the room contains eight conference chairs, as well as a big meeting table, a computer desk, and a bookshelf of the sizes (2.0, 4.0, 0.7), (0.6, 1.9, 0.7), and (1.5, 0.5, 1.8)\,m, respectively. In Fig.\,\ref{fig:furniture}(c), the meeting table and chairs do not vertically occlude the audio device from the floor. 
In this scenario, unlike Figs.\,\ref{fig:furniture}(a) and (b), the area occupied by the bookshelf is excluded from the predicted floorplan map. The difference between the bookshelf and other objects is their height. The heights of the meeting table, chairs, and computer desk are lower than the vertical position of the audio device, so those objects did not significantly alter the geometry of the estimated room. In contrast, the bookshelf with a large vertical dimension is recognized as a wall by the model. Another effect of occlusion can be observed in Fig.\,\ref{fig:furniture}(d), where the audio device is placed 0.5\,m above the meeting table.
In this case, the meeting table occludes the direct sound wave propagating from the audio device to the floor, resulting in an erroneous prediction of the floor position. The predicted floor position is 0.5\,m below the device in the estimated height map and corresponds to the position of the meeting table. However, the floorplan map is not significantly altered compared to Fig.\,\ref{fig:furniture}(c). 
The experiments shown here only deal with a limited number of cases, and more extensive studies are required to investigate the general behavior of EchoScan against various indoor objects. In addition, some of these degradations may be reduced by fine-tuning or additional training through more diverse RIR datasets simulated with indoor objects. Nevertheless, this limited case study demonstrates that (1) EchoScan is not completely collapsed by small indoor objects and has some robustness, and (2) large furniture with a size comparable to the wall dimension is treated like a wall and reduces the size of the estimated room.}

\subsection{{Influence of Violated Conditions}}
{For training EchoScan, we introduced \rev{four} conditions: (1) room size parameters defining room area within the floorplan and height maps have limited ranges, (2) the audio device is not too close to the walls, \rev{(3) the absorption or scattering coefficients of the walls are not extremely high, and (4) the loudspeaker is omnidirectional.} In this subsection, we investigate the capabilities of EchoScan when these conditions are violated.}

{First, we examine the case where the room sizes are larger than those used for the training data. To this end, we created room layouts exceeding the maximum size of the basic room dataset described in Section\,\ref{sec:Basic Room Dataset}. The original size parameters of the basic room dataset were chosen within the ranges $[2, 5]$, $[2, 5]$, and $[3, 5]$\,m for length, width, and height, respectively, yielding the maximum side length of a 2D polygon of approximately 10\,m. For this experiment, we generated a quadrilateral room with dimensions $(13, 8, 4)$\,m, exceeding the basic room dataset's maximum side length limit. The acoustic materials used in this experiment are as follows: hard surface for the sidewalls; linoleum on concrete for the floor; gypsum board for the ceiling.}

\begin{figure}[t]
    \centering
    \includegraphics[width=0.9\columnwidth]{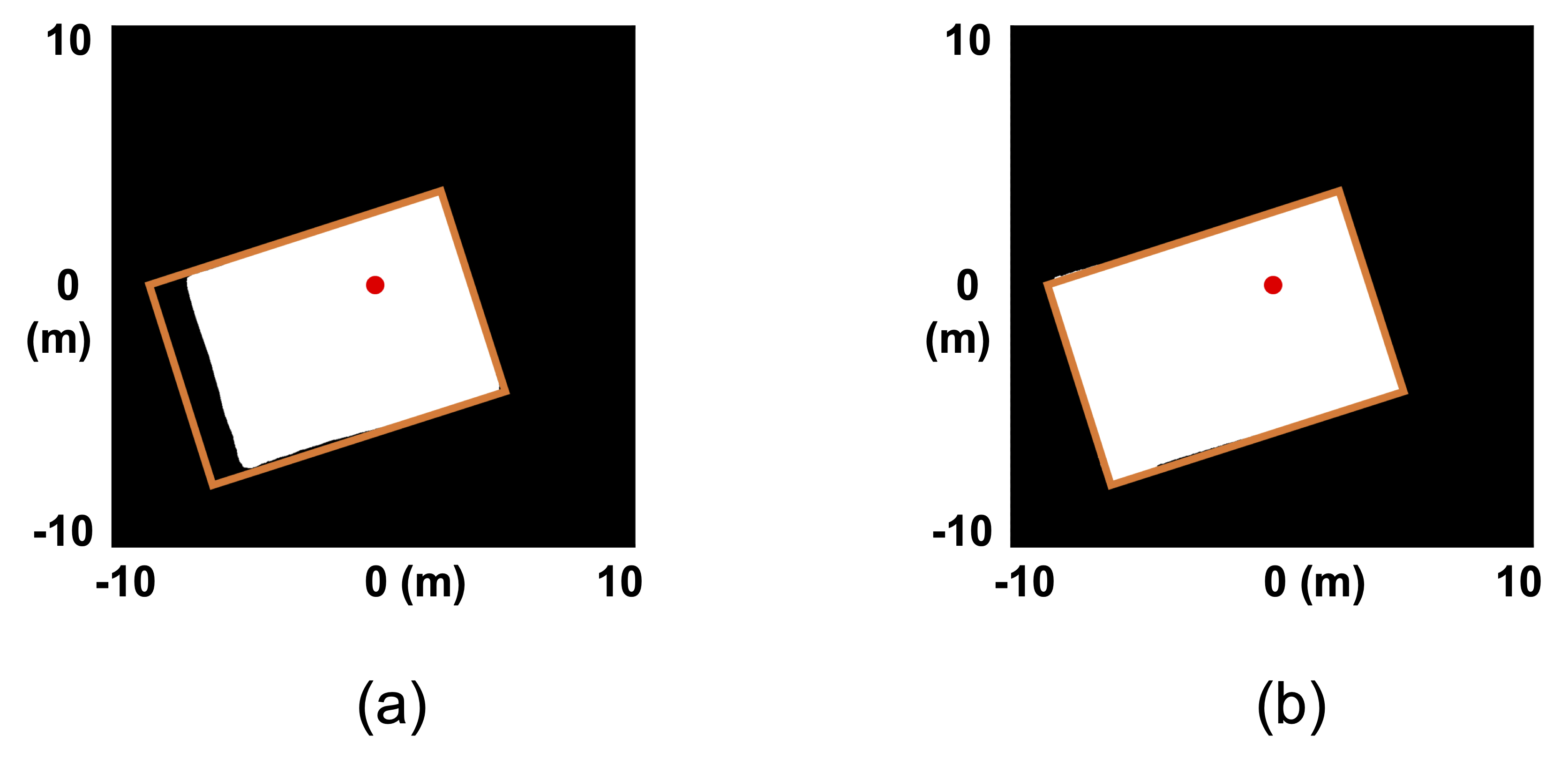}
    \caption{{EchoScan's extrapolation capability study for a room larger than the ranges of the room size parameters of the basic room dataset. The thick orange line represents the boundaries of the GT room, and the red dot indicates the device position. The GT room has dimensions of (13, 8, 3.5)\,m, whose longer side exceeds the maximum length (10\,m) of the basic room dataset. (a) Inference result of the EchoScan model trained only using the basic room dataset. (b) The inference result of the EchoScan model fine-tuned using a Manhattan-Atlanta room dataset. During fine-tuning, room size variations within the range of [0.5, 2] were utilized to augment the limited data in the Manhattan-Atlanta room dataset.}}
    \label{fig:extrapolation_large}
\end{figure}

\begin{figure}[t]
    \centering
    \includegraphics[width=\columnwidth]{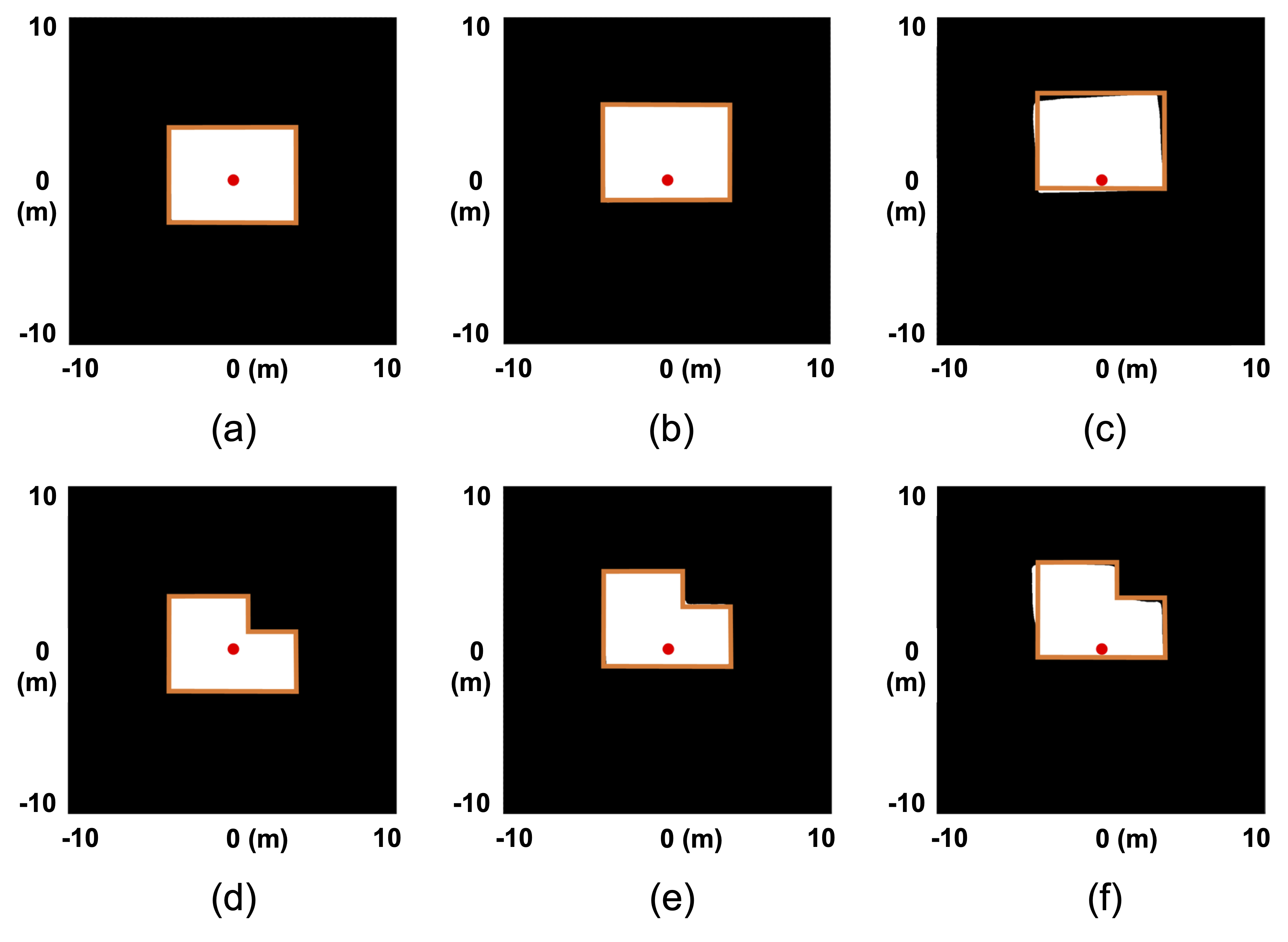}
    \caption{{EchoScan's extrapolation capability study when the audio device is located outside of the 70\% area of a given room. The thick orange line represents the boundaries of the GT room, and the red dot indicates the device position. These results imply that EchoScan can accurately predict room geometry even when the audio device is located outside of the 70\% area of the given rooms, which is an unseen range during training. (a), (b), and (c) show the device located 2.5\,m, 1\,m, and 0.4\,m away from the bottom side of a quadrilateral GT floorplan map, respectively. (d), (e), and (f) show the device located 2.5\,m, 1\,m, and 0.4\,m away from the bottom side of an L-shaped GT floorplan map, respectively.}}
    \label{fig:extrapolation_device}
\end{figure}

{Fig.\,\ref{fig:extrapolation_large}(a) shows the inference result of EchoScan trained solely on the basic room dataset. The inferred quadrilateral floorplan map measures approximately $(11, 8)$\,m, which is greater than the side length limit but less than the GT length of 13\,m. From this result, we can see that EchoScan's inference is influenced by the size limit of the training data. However, this problem was resolved when we fine-tuned EchoScan using the Manhattan-Atlanta room dataset.
Fig.\,\ref{fig:extrapolation_large}(b) demonstrates that the fine-tuned EchoScan can estimate the oversized quadrilateral room without a problem. This is because the random room scaling within the scale factor of range $[0.5, 2]$ was applied to augment the limited number of Manhattan-Atlanta room datasets. These results stress that the limited room size of the train dataset does impact the predictable room size of EchoScan but the size issue can be resolved by training the model with room layouts of appropriate sizes.}

{The second experiment is for the audio device positioned close to the walls. In Section\,\ref{sec:problem}, we assume that the audio device can be placed within 70\% of the length-width space of the given room, but in this experiment, the device was gradually moved towards one of the walls to violate the assumption. The acoustic materials used in this experiment are as follows: hard surface for the sidewalls; linoleum on concrete for the floor; gypsum board for the ceiling.} 

{Fig.\,\ref{fig:extrapolation_device} illustrates the GT walls (bold orange lines) and inferred floorplan maps corresponding to different device positions (red dots) in quadrilateral and L-shaped rooms. Figs.\,\ref{fig:extrapolation_device}(a), (b), (c) show the audio device located approximately 2.5\,m, 1\,m, 0.4\,m from the wall at the bottom side of the map, respectively, in a quadrilateral room constructed by the room size parameters of $\bs=[4, 3, 4]\T$\,m, which has dimensions of (8, 6, 4)\,m. Figs.\,\ref{fig:extrapolation_device}(d)--(f) show the inference results for the same situation in the L-shaped room constructed by the room size parameters of $\bs=[4, 3, 4]\T$\,m and cutout positions of $\bm{\mu}^L=[1, 0.75]\T$\,m. The shortened distance to the wall only slightly affected the estimated room shape, even in the case of Figs.\,\ref{fig:extrapolation_device}(c) and (f), where the 70\% length-width space assumption is broken.
Still, a more extensive analysis with various device positions is necessary but these case studies show the possibility of estimating room layouts for device positions unseen during the training.}

\begin{figure}[t]
    \centering
    \includegraphics[width=0.93\columnwidth]{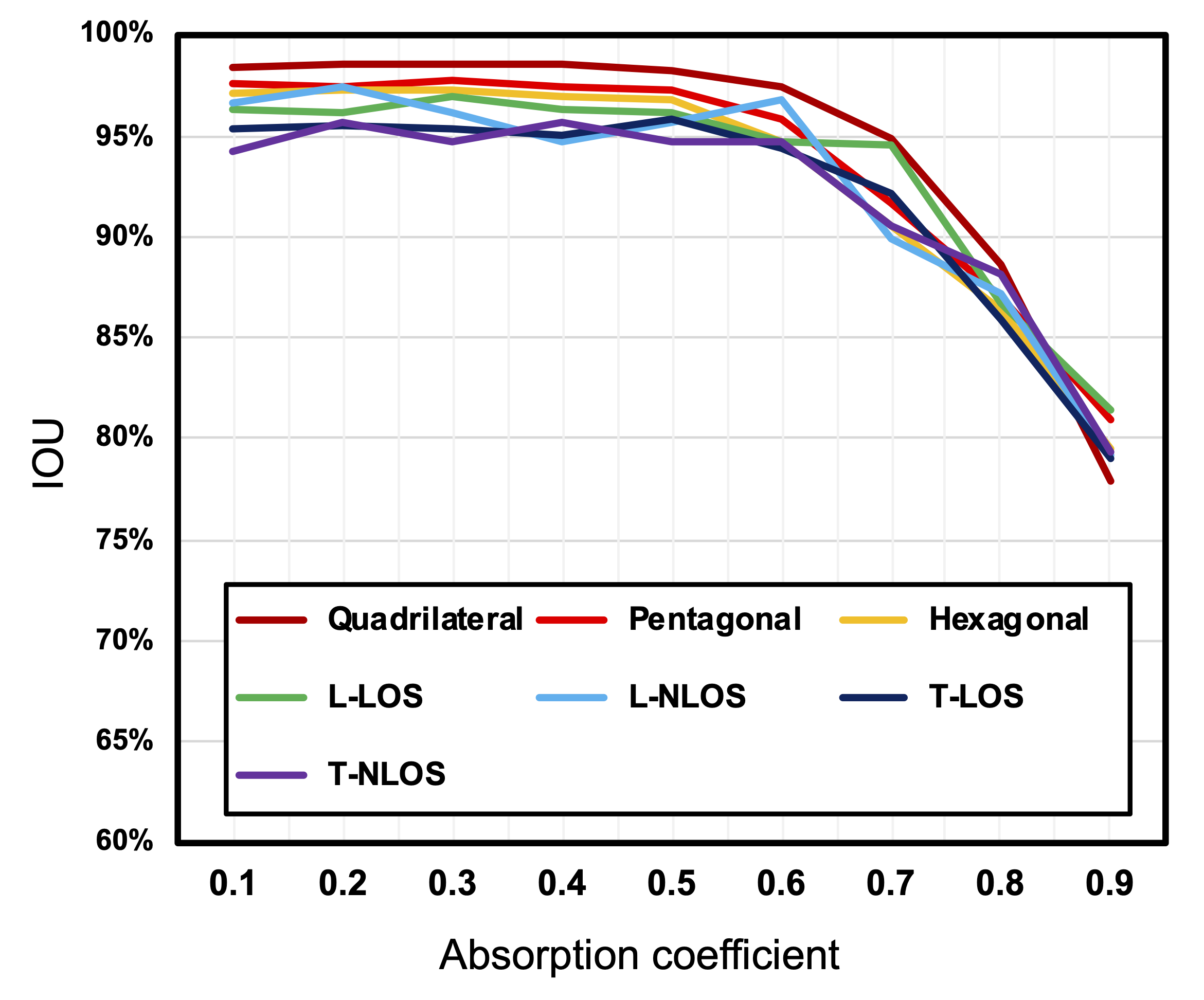}
    \caption{{Influence of absorption coefficients on RGI performance (IOU). The same absorption coefficient was applied to all walls and across all frequency bands. The scattering coefficient was fixed to 0.1.}}
    \label{fig:abs_coeff}
\end{figure}

\begin{figure}[t]
    \centering
    \includegraphics[width=0.95\columnwidth]{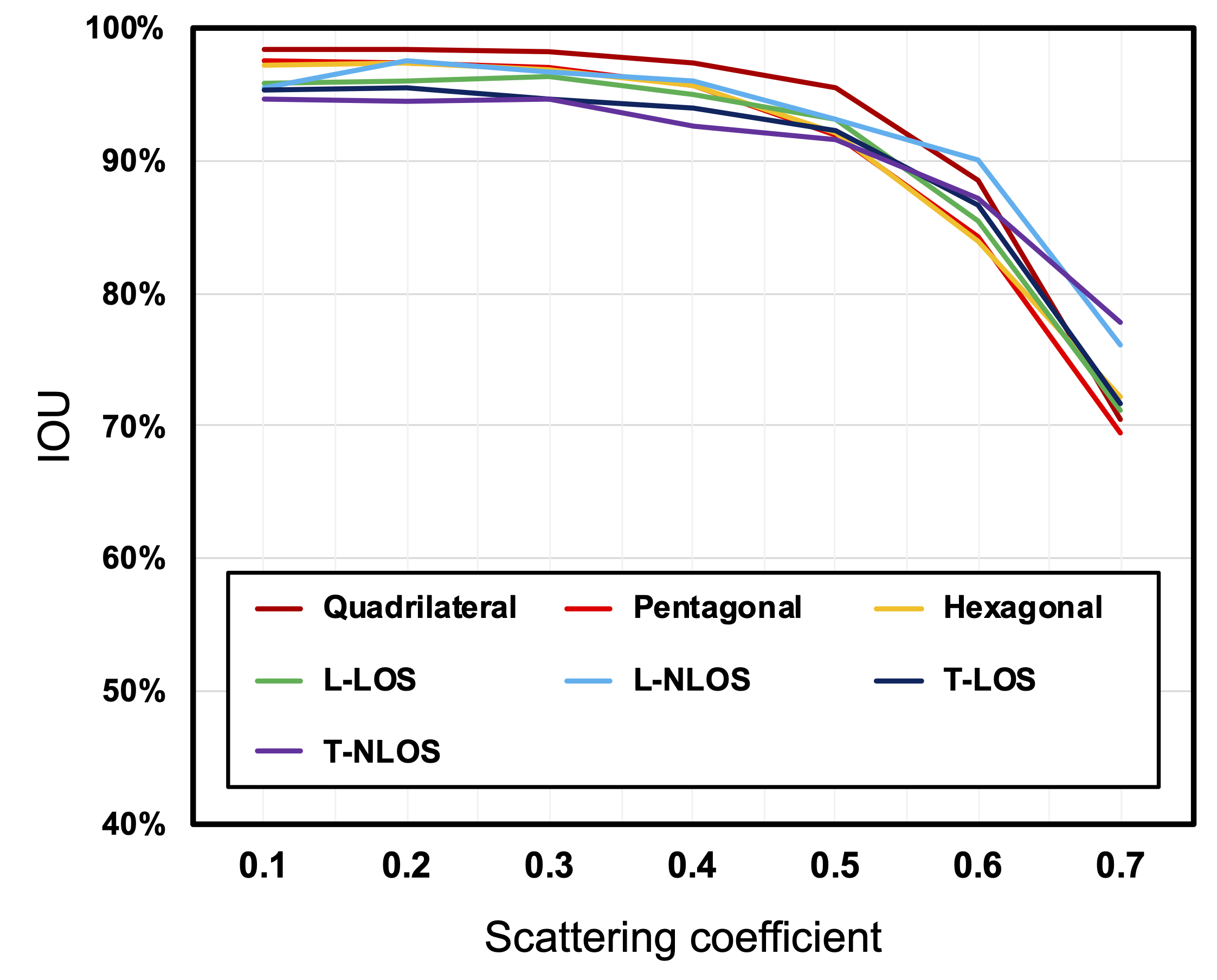}
    \caption{{Influence of scattering coefficient on RGI performance (IOU). The same scattering coefficient was applied to all walls and across all frequency bands. The absorption coefficient was fixed to 0.1.}}
    \label{fig:scatter_coeff}
\end{figure}

The third experiment involves materials with high absorption coefficients or scattering coefficients. During training, EchoScan is exposed to various acoustic materials such as gypsum boards, plasterboards, carpet, concrete, and wooden materials, typically used for indoor spaces. Despite the training with diverse materials, the RGI performance can decrease when absorption or scattering coefficients are very high. Fig.\,\ref{fig:abs_coeff} demonstrates the change in the performance of EchoScan with respect to the absorption coefficients. For this experiment, the same absorption coefficient was applied to all walls of rooms in the basic room dataset and across all frequency bands, while the scattering coefficient was fixed to 0.1. EchoScan remains robust until the absorption coefficient reaches 0.6, but its performance decreases rapidly from 0.7. With such a high absorption coefficient, early reflections quickly lose their energy after several reflections. Strong scattering from walls can be another problematic factor. Fig.\,\ref{fig:scatter_coeff} presents the IOU change according to the increase of the scattering coefficient with the absorption coefficient fixed to 0.1. EchoScan robustly infers room geometries for the scattering coefficient less than 0.6 but its performance is reduced for high scattering coefficients exceeding 0.6. Therefore, strong diffuse reflections spreading out reflection peaks can limit the RGI ability of EchoScan. These case studies show that EchoScan may struggle for walls made up of such a highly absorbing or scattering material.

\begin{figure*}[t]
    \centering
    \includegraphics[width=\textwidth]{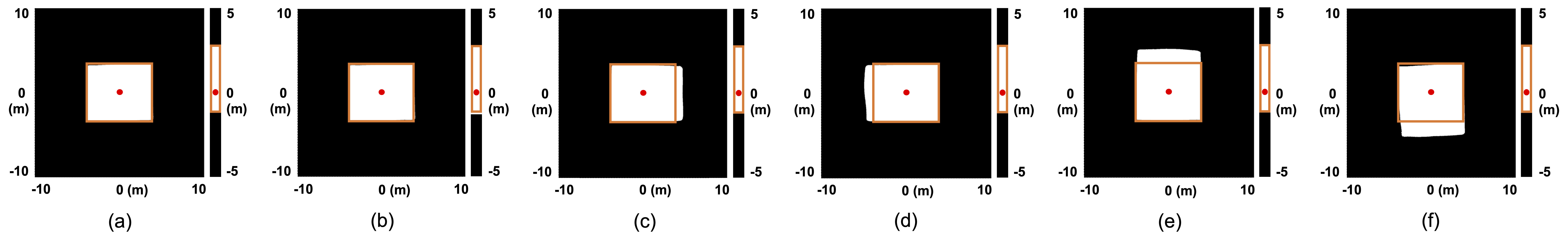}
    \caption{\rev{Influence of directive loudspeaker on the estimated floorplan map and height map. The thick orange line represents the boundaries of the GT room, and the red dot indicates the device position. (a) The loudspeaker is omnidirectional. (b)--(f) The on-axis of the directive loudspeaker faces the positive height axis, the negative length axis, the positive length axis, the negative width axis, and the positive width axis, respectively.}}
    \label{fig:directivity}
\end{figure*}

\rev{The fourth experiment presents RGI results with directive loudspeakers (cardioid directivity). In this experiment, a quadrilateral room with dimensions $(8, 7, 4)$\,m was considered, with absorption coefficients set to 0.1. The RIRs were simulated using the image source method up to the sixth-order reflections, as Pyroomacoustics \cite{pyroomacoustics} supports the simulation of source directivity only for the image source method. Fig.\,\ref{fig:directivity} shows the GT walls (bold orange lines) and the inferred floorplan maps and height maps corresponding to the different orientations of the cardioid loudspeaker. Fig.\,\ref{fig:directivity}(a) displays the case where the loudspeaker is omnidirectional, while Figs.\,\ref{fig:directivity}(b)--(f) depict cases where the on-axis of the cardioid loudspeaker is oriented toward the positive height axis, the negative length axis, the positive length axis, the negative width axis, and the positive width axis, respectively. In Fig.\,\ref{fig:directivity}(b), when the on-axis of the loudspeaker’s directivity faces the ceiling, there is no significant degradation in the estimated floorplan but the estimated height deviates slightly. In Figs.\,\ref{fig:directivity}(c)--(f), on the other hand, when the on-axis of the loudspeaker’s directivity faces the sidewalls, an overestimated area appears in the off-axis direction, i.e., the null direction of the cardioid pattern. These results indicate that the decreased reflections due to the source directivity degrades the RGI accuracy of EchoScan.}

\section{{Challenges and Limitations}}
{In this study, we aimed to infer room geometry using an audio device similar to off-the-shelf voice assistant speakers. Despite the ability of EchoScan demonstrated in complex and diverse room geometries, several challenges remain for real-world scenarios.}

{The first challenge involves the discrepancy between simulated and real sound propagation. The acoustic simulation tools we utilized cannot perfectly mimic the propagation of real sound waves, leading to potential performance reduction in real-world scenarios. In particular, Pyroomacoustics uses the raytracing or image source method for acoustic simulation, but raytracing has limitations in simulating low-frequency sound fields, diffractions by room corners and indoor objects, and transmission through partitions and walls.}

{The second challenge concerns discrepancies in loudspeakers and microphones. We assume transparent sound radiation from a point source (omnidirectional). However, real loudspeakers and microphones have frequency-dependent directivities and self-scattering by the enclosure of an audio device itself. Off-the-shelf voice assistant speakers have various acoustical designs for the loudspeaker, and differences in frequency responses exist across manufacturers and models. While some of these discrepancies might be addressed through fine-tuning with RIRs measured from real audio devices, constructing fine-tuning datasets for various rooms is resource-intensive. Therefore, a generalized DNN model that operates without fine-tuning would be preferred.}

{Lastly, EchoScan is built on the \rev{geometrical} assumption that a room consists of a parallel floor and ceiling combined with side walls of finite size, \rev{and the acoustical assumption that the absorption or scattering coefficients of the walls are not excessively high. In real-world indoor spaces, however, rooms may have non-parallel floors and ceilings or may contain walls made of acoustic materials with high absorption or scattering coefficients across all frequencies, such as thick polyurethane foam (high absorption coefficient), thick mineral wool (high absorption coefficient), or acoustic diffuser panels (high scattering coefficient).} To apply EchoScan to various real-world indoor spaces, a DNN model capable of inferring room geometry without these assumptions needs to be developed.}

{Despite these remaining challenges, EchoScan is the first acoustic-based RGI model capable of inferring Atlanta and Manhattan layouts using a single audio device positioned at a single position. The case studies also demonstrate that the model has some resilience to parameter changes unseen during training. The model's ability to detect NLOS walls can be even more valuable when combined with the vision-based method, overcoming the modality-specific weaknesses through a multimodal approach.}

\section{Conclusion}
In this study, we introduced EchoScan, a pioneering deep neural network model that utilizes high-order acoustic echoes to infer NLOS walls and complex-shaped rooms using RIRs measured by a single voice assistant speaker. EchoScan is a pixel-segmentation network that infers room geometry as a combination of 2D floorplan and 1D height maps, enabling the representation of various room shapes that are difficult to express using traditional wall equations. Our model employs an encoder-decoder structure to generate these floorplan and height maps by comprehensively understanding RIR data, including high-order reflections.

The RGI performance of EchoScan was validated using both the basic room dataset with simple room layouts and the Manhattan-Atlanta room dataset including Manhattan and Atlanta layouts. The RGI results from both datasets demonstrated the robustness and generalization ability of the EchoScan across diverse room geometries. 
We also confirmed that EchoScan utilizes information from higher-order reflections to infer complex room geometries. Grad-CAM activation maps showed that the model emphasizes high-order reflections when predicting more complex room geometries. Ablation studies conducted with truncated RIRs also revealed significant performance degradation in the absence of high-order reflections. 
Further generalization studies, carried out with indoor objects and audio devices closely positioned to the walls, underscore the robustness of EchoScan against unexpected perturbations in the experimental setting. These results demonstrate the potential of EchoScan as the acoustic-based foundation model for RGI tasks, resolving the limitations of previous models on curved and NLOS walls.

\rev{The remaining challenges for EchoScan arise from its intrinsic assumptions, including the maximum allowable room size, the absence of large occluding objects, and the requirement for parallel floors and ceilings. Another key issue is the generalization to various directivities and frequency responses of loudspeakers and microphones, as well as diverse wall materials. Addressing these limitations is a key future direction for developing a more general RGI model that can adapt to a wider range of room environments.}

\bibliographystyle{Bibliography/IEEEtran}
\bibliography{Bibliography/references}

\begin{IEEEbiography}[{\includegraphics[width=1in,height=1.25in,clip,keepaspectratio]{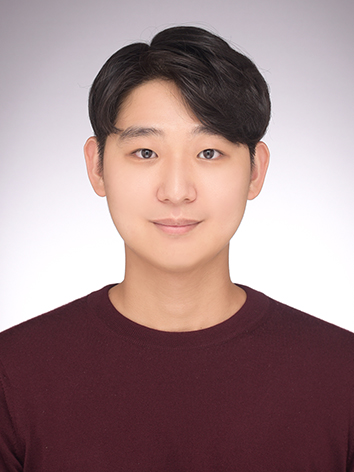}}]
{Inmo Yeon} (Student Member, IEEE) received an A.A. degree in audio production from the Dong-ah Institute of Media and Arts (DIMA), South Korea in 2018; a B.S. degree in audio engineering from the National Institute for Lifelong Education (NILE), South Korea in 2019; and an M.S. degree in electrical engineering from Hanyang University, South Korea in 2021. He is currently pursuing a Ph.D. degree at the School of Electrical Engineering of the Korea Advanced Institute of Science and Technology (KAIST), South Korea. His research interests include signal processing, room acoustics, spatial audio, and deep learning.
\end{IEEEbiography}

\begin{IEEEbiography}[{\includegraphics[width=1in,height=1.25in,clip,keepaspectratio]{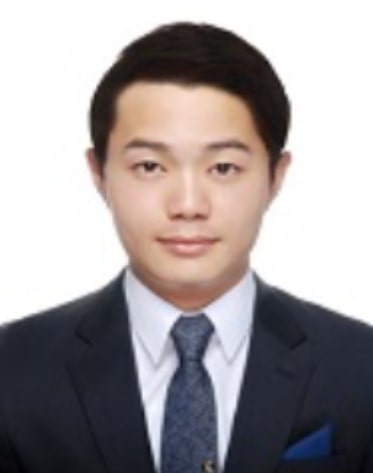}}]
{Iljoo Jeong} received a B.S. degree in mechanical engineering from Ajou University, Suwon, South Korea in 2016. He is currently a Ph.D. candidate in the Department of Mechanical Engineering at Pohang University of Science and Technology (POSTECH), South Korea. He is currently a guest researcher at the
Korea Research Institute of Standards and Science (KRISS), Daejeon, South Korea. His research interests include deep learning, smart manufacturing, acoustics, and inverse problem-solving.
\end{IEEEbiography}

\begin{IEEEbiography}[{\includegraphics[width=1in,height=1.25in,clip,keepaspectratio]{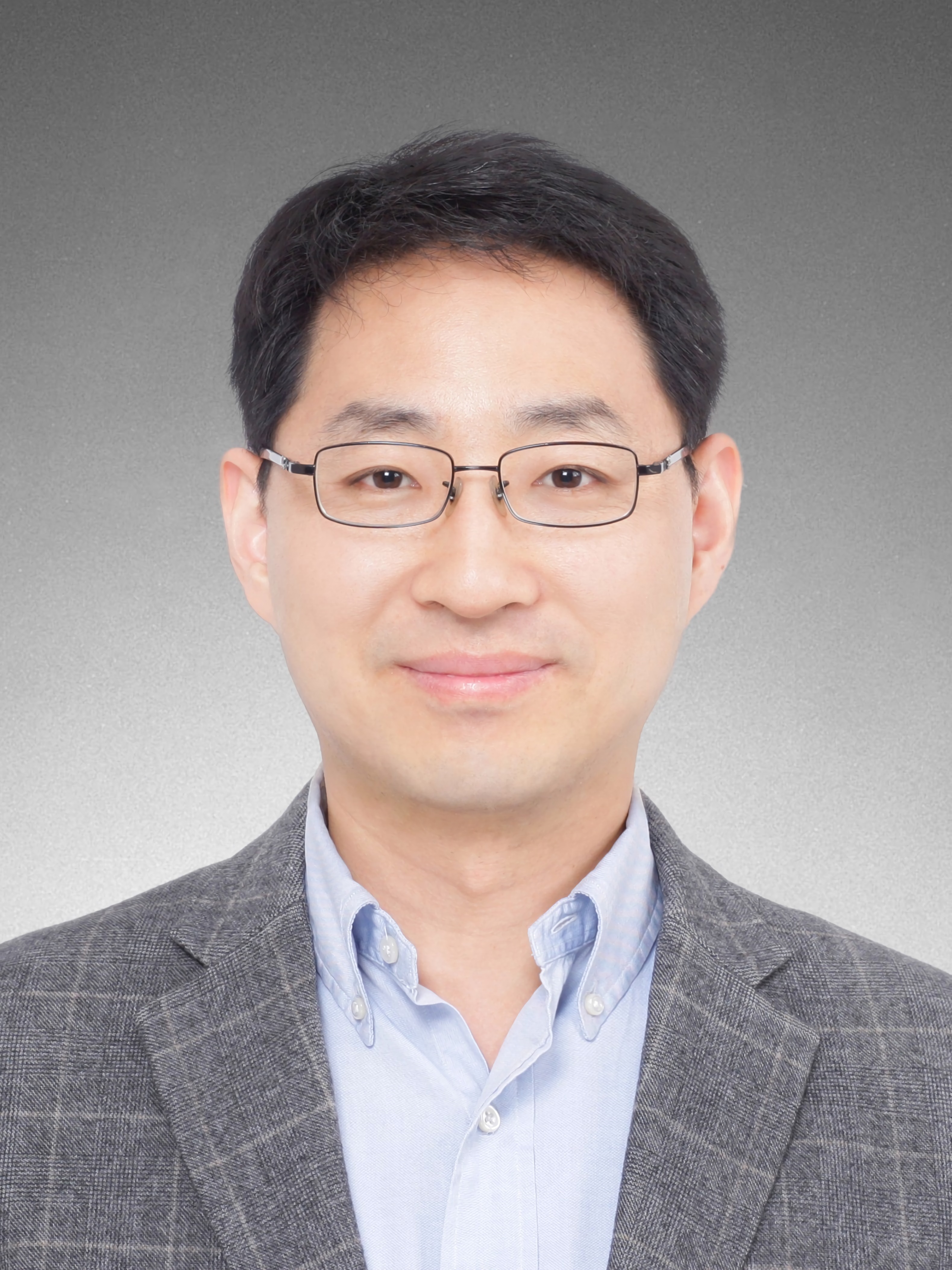}}]
{Seungchul Lee} received a B.S. degree in mechanical and aerospace engineering from Seoul National University, Seoul, South Korea in 2001, and M.S. and Ph.D. degrees in mechanical engineering from the University of Michigan, Ann Arbor, MI, USA in 2008 and 2010, respectively. He is currently an Associate Professor at the Department of Mechanical Engineering of the Korea Advanced Institute of Science and Technology (KAIST), Daejeon, South Korea. His research focuses on industrial artificial intelligence for mechanical systems, smart manufacturing, materials, and healthcare. His research extends to both knowledge-guided AI and AI-driven knowledge discovery.

\end{IEEEbiography}

\begin{IEEEbiography}[{\includegraphics[width=1in,height=1.25in,clip,keepaspectratio]{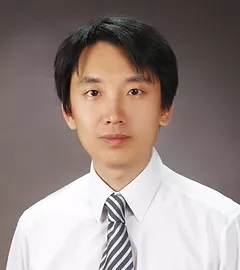}}]
{Jung-Woo Choi} (Member, IEEE) received the B.Sc., M.Sc., and Ph.D. degrees in mechanical engineering from the Korea Advanced Institute of Science and Technology (KAIST), South Korea in 1999, 2001, and 2005, respectively. From 2006 to 2007, he was a postdoctoral researcher at the Institute of Sound and Vibration Research of the University of Southampton, Southampton, U.K. From 2007 to 2011, he worked with Samsung Electronics at the Samsung Advanced Institute of Technology, Suwon, South Korea. He was a Research Associate Professor in the Department of Mechanical Engineering at KAIST until 2014. In 2015, he joined the School of Electrical Engineering of the KAIST as an Assistant Professor. In 2018, he became an Associate Professor. His current research interests include sound-field reproduction, sound focusing, array signal processing, and their applications. He is a member of the Acoustical Society of America at the Institute of Noise Control Engineers, USA, and the Korean Society of Noise and Vibration Engineering.
\end{IEEEbiography}

\end{document}